\documentclass[superscriptaddress,aps,pra,nofootinbib,onecolumn,twoside,reprint,letterpaper,showpacs,10pt,floatfix]{revtex4-2}

\usepackage{graphicx}% Include figure files
\usepackage{dcolumn}% Align table columns on decimal point
\usepackage{bm}% bold math
\usepackage{amsmath}
\usepackage{xcolor}
\usepackage{hyperref}
\usepackage{tablefootnote}
\usepackage{verbatim} 

\usepackage{caption}
\usepackage{subfig}

\newcommand{\gagg}{g_{a\gamma\gamma}}
              
\begin{document}

\title{Maximizing Quantum Enhancement in Axion Dark Matter Experiments}

\author{Chao-Lin Kuo}
\affiliation{%
 Stanford University, Stanford, CA 94305, USA
}%
\affiliation{%
 SLAC National Accelerator Laboratory, 2575 Sand Hill Road,
Menlo Park, CA 94025, USA
}%

\author{Chelsea L. Bartram}
\affiliation{%
 SLAC National Accelerator Laboratory, 2575 Sand Hill Road,
Menlo Park, CA 94025, USA
}%

\author{Aaron S. Chou}
\affiliation{%
 Fermi National Accelerator Laboratory, Batavia, IL 60510, USA
}%

\author{Taj A. Dyson}
\affiliation{%
 Stanford University, Stanford, CA 94305, USA
}%

\author{Noah A. Kurinsky}
\affiliation{%
 SLAC National Accelerator Laboratory, 2575 Sand Hill Road,
Menlo Park, CA 94025, USA
}%

\author{Gray Rybka}
\affiliation{%
University of Washington, 1410 NE Campus Parkway
Seattle, WA 98195, USA
}%

\author{Sephora Ruppert}
\affiliation{%
 Stanford University, Stanford, CA 94305, USA
}%

\author{Osmond Wen}
\affiliation{%
 Stanford University, Stanford, CA 94305, USA
}%
\affiliation{%
 California Institute of Technology, 1200 E California Blvd, Pasadena, CA 91125, USA
}%

\author{Matthew O. Withers}
\affiliation{%
 Stanford University, Stanford, CA 94305, USA
}%

\author{Andrew K. Yi}
\affiliation{%
 SLAC National Accelerator Laboratory, 2575 Sand Hill Road,
Menlo Park, CA 94025, USA
}%

\author{Cheng Zhang}
\affiliation{%
 Stanford University, Stanford, CA 94305, USA
}%

\date{\today}% It is always \today, today,

\begin{abstract}
\noindent

We provide a comprehensive comparison of linear amplifiers and microwave photon-counters in axion dark matter experiments.  The study is done assuming a range of realistic operating conditions and detector parameters, over the frequency range between 1--30 GHz. 
As expected, photon counters are found to be advantageous under low background, at high frequencies ($\nu>$ 5 GHz), {\em if} they can be implemented with robust wide-frequency tuning or a very low dark count rate.
Additional noteworthy observations emerging from this study include: (1) an expanded applicability of off-resonance photon background reduction, including the single-quadrature state squeezing, for scan rate enhancements; (2) a much broader appeal for operating the haloscope resonators in the over-coupling regime, up to $\beta\sim 10$; (3) the need for a detailed investigation into the cryogenic and electromagnetic conditions inside haloscope cavities to lower the photon temperature for future experiments; (4) the necessity to develop a distributed network of coupling ports in high-volume axion haloscopes to utilize these potential gains in the scan rate.
\end{abstract}
\maketitle

\section{Introduction}

\noindent 

QCD axions are excellent candidates for the cosmological dark matter  \cite{PecciQuinn1977,Weinberg1978,Wilczek1978,abbott83,marsh}. Their presence in a terrestrial laboratory can be detected using a well-established method, the cavity-based haloscope \cite{sikivie1,admx18,haystac18,capp}.  In this scheme, a cavity resonator is immersed in an external dc magnetic field, resonantly converting halo axions into a microwave signal. The resonant frequency of the cavity can be tuned to search for a narrow line which would indicate the presence of axion dark matter. The cavity is coupled to a low-noise amplifier for readout and data processing. 

In recent years, advances in quantum sensing techniques enable new capabilities in axion dark matter searches\cite{lamoreaux,sushkov}. It is now possible to apply vacuum squeezing techniques to evade the standard quantum limit (SQL) by using a pair of parametric amplifiers operating in a single-quadrature squeezed state setup \cite{malnou,backes,jewell}.  It has also been demonstrated that a microwave photon counter can be used in cm-wave axion dark matter searches \cite{lescanne,albertinale,balembois,braggio}, which enhances the scan rate by avoiding SQL altogether.  

In this paper, we provide an in-depth comparison of the linear amplifiers and the photon-counting microwave detectors in axion dark matter searches. Revisiting this subject is timely given these advances in quantum information technologies.  Obviously, such promises can only be realized when auxiliary systems in the haloscope experiments can support operation of these new sensors under appropriate conditions. The purpose of this review is to clarify what these requirements are and to develop a road map for future axion dark matter searches in the frequency range of 1 -- 30 GHz, corresponding to the post-inflationary scenario of axion production \cite{marsh,Borsanyi2016,Klaer_2017,Graham_2018,Takahashi_2018,Buschmann2022}. 

%To broaden the applicability of these results to include ongoing experiments, we consider not only the zero-background case, but also low-background when the photon occupation number $n$ is of order unity, and high-background when $n$ is of order 10 and above. A well-known result in axion dark matter research is the scan rate of a resonator haloscope is maximized when the coupling coefficient, commonly called the $\beta$ parameter, is equal to two. This result is obtained under the assumption of $\geq$ SQL phase insensitive amplification.  It is also known that when the vacuum squeezing technique is used, the scan rate is optimized when $\beta$ is around five.  
%In this work, we revisit the optimization of the coupling strength in an expanded variety of experimental configurations. We consider the case of cooling the termination resistor that provides the background photon field to a much lower temperature than the temperature of the haloscope. We also consider the case of reading out the haloscope with a microwave photon counter instead of an amplifier.

The paper is organized as follows. In \ref{sec:scanrate} we derive the scan rate of an axion haloscope read out by either a linear amplifier (with the option of adding a squeezing de-amplifier) or a single microwave photon detector. We obtain closed-form expressions of which we take various limits -- photon noise background- versus dark count rate (DCR)- limits, as functions of the temperature and search frequencies.  In \ref{sec:compare}, we proceed to use the results to comprehensively compare these two readout systems, and their ability to reach theoretical benchmarks such as the KSVZ \cite{kim,shifman} and DFSZ \cite{Zhitnitsky:1980tq,DINE1981199}.  We conclude in \ref{sec:conclude} and discuss the implications of these results on future searches of the post-inflationary dark matter axions.

\begin{figure}[b]
    \centering
    \includegraphics[width=0.6 \columnwidth]{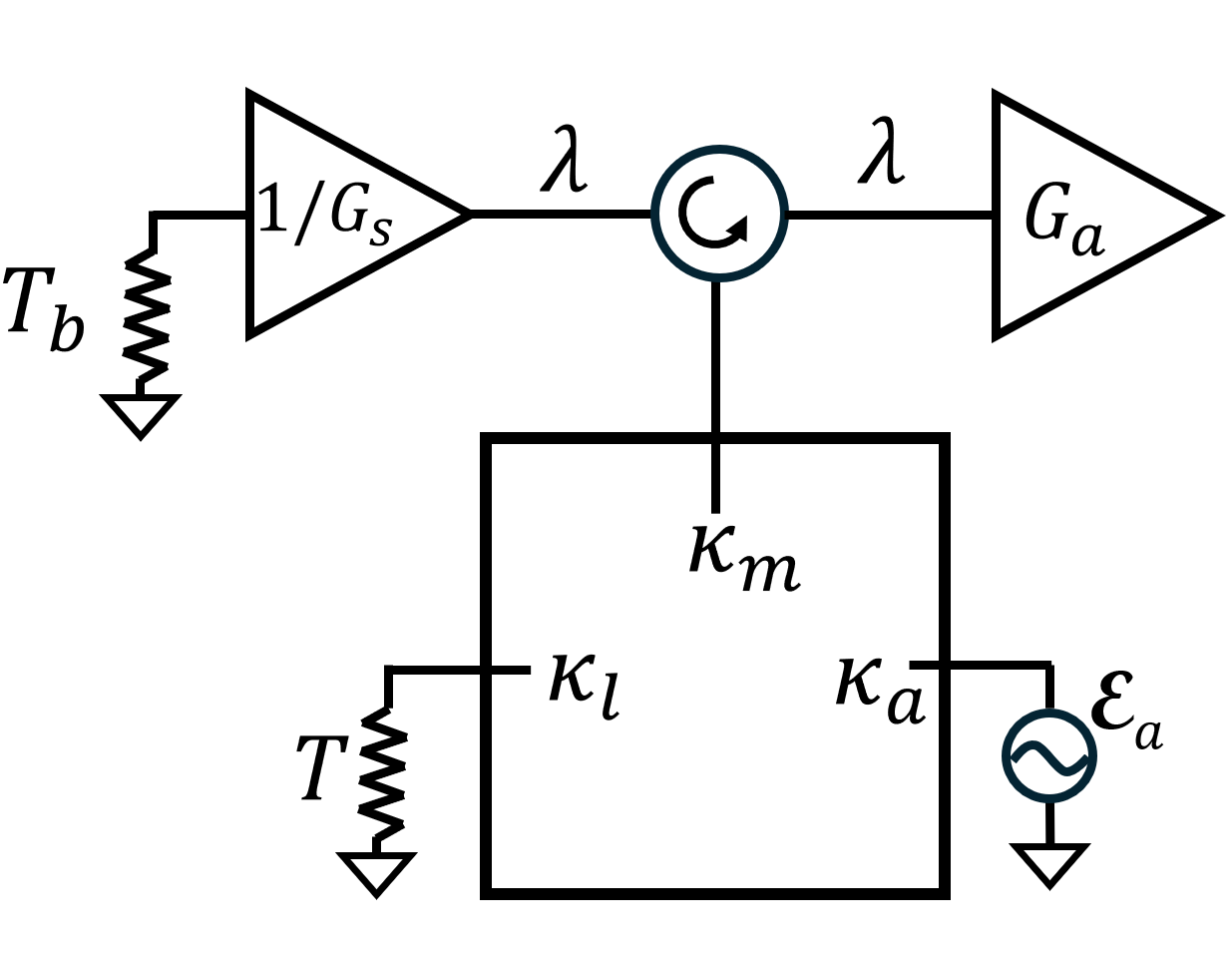}
    \caption{The experimental setup considered in \S 2A and 2B. The schematic and notation closely follows those of \cite{malnou}. $G_a$ and $G_s$ are the amplifier and squeezer gains. The coupling coefficient $\kappa_m$, $\kappa_l$, and $\kappa_a$ are for the measurement port $m$, intrinsic (cavity) loss port $l$, and the axion port $a$.  In this paper, we consider the general case where $T_b$, the termination temperature, is different from the haloscope temperature $T$. }
    \label{fig:amp}
\end{figure}

\section{The Scan Rate Calculations}
\label{sec:scanrate}

To calculate the scan rate $R$ of the halsocope for reaching a certain coupling strength $\gagg$, we closely follow the method and notations adopted in a well cited paper, reference \cite{malnou}. We consider the experimental setup depicted in Fig. \ref{fig:amp}. The large square block represents the cavity haloscope, which is coupled to a measuring port $m$, an internal lossy termination (surface conductive loss) at the haloscope temperature $T$, and a generator $a$ from axion-to-photon conversion.  The coupling strengths are, respectively, $\kappa_m$, $\kappa_l$, and $\kappa_a$. 

The measurement port $m$ is connected to a circulator whose output goes to a low noise amplifier to the right and the input from an impedance matching termination to the left, with the efficiency $\lambda\leq 1$ in either direction. Optionally, a quadrature de-amplifier with a de-amplification gain $G_s$ can be inserted between the circulator and the termination. We include loss in efficiency due to dissipation, with $\lambda$ being the efficiency to either the amplifier side or the termination side.

While we closely follow \cite{malnou}, we consider the general case where the termination temperature $T_b$ can be different, usually lower, than the haloscope temperature $T$. This situation arises in axion experiments because the large surface area of the haloscope cavity and the poorly thermal anchored tuning elements often make it difficult to operate the haloscope at the base temperature $T_b$ of the mixing chamber \cite{backes,jewell}. On the other hand, the small coaxial termination can be much more easily heat sunk to the base temperature $T_b$ of the refrigerator.  

Starting with the Heisenberg-Langevin equation for the cavity ladder operator $\hat A$ and following the steps based on known solutions to the input-output theory as presented in Appendix A of \cite{malnou}, we obtain the single-quadrature spectral density at the amplifier:

%\begin{widetext}

\begin{multline}
\label{sigma}
\mathrm{\Sigma}_{out,X,m}=(n_T+\frac{1}{2})\ (1-\lambda)+\frac{\lambda}{B(\omega)}\times\\
\Biggl[(n_A+\frac{1}{2}) \kappa_a \kappa_m+(n_T+\frac{1}{2}) \kappa_l \kappa_m\\
+(n_b+\frac{1}{2})(1-\lambda+\lambda/G_s )A(\omega)\Biggr],
\end{multline}

%\end{widetext}

where $A(\omega)\equiv\omega^2+(\kappa_m-\kappa_l)^2/4$ ; $B(\omega)\equiv\omega^2+(\kappa_m+\kappa_l)^2/4 \ \ $; and the photon occupation number $n$ for thermal sources is given by $(e^{\hbar \omega/kT}-1)^{-1}$.  The coefficients $A(\omega)$ are simply the reflectivity of the resonator $|\Gamma|^2$ in classical microwave theory and $B(\omega)$ a normalization factor.  The coupling constants $\kappa_m,\kappa_l$ can be connected with their microwave equivalents by the following identifications: $\kappa_l/\omega_c=\kappa_l/(2\pi \nu_c)=1/Q_0$ and $\kappa_m=\beta\kappa_l$, where $Q_0$ is the intrinsic (unloaded) quality factor of the resonator and $\beta$ is the coupling parameter of the measurement port. This setup can be analyzed as a cascaded microwave network. At the resonant frequency of the cavity, the amplifier ``sees'' thermal emission from the cavity (especially if $\beta=1$). When the cavity becomes reflective coming off resonance, the amplifier sees the radiation from the termination. 

The photon occupation number $n_A$ ($\gg 1$) and coupling $\kappa_a$ can be expressed in terms of physical parameters normally found in the haloscope literature, such as $\gagg$ and the form factor \cite{malnou}.  We reproduce their Eqns B(10) \& B(11) here

\begin{equation}
\label{na}
n_A=\frac{\lvert\gagg\rvert\rho_a B_0V}{4\omega_a\Delta_a}\sqrt{\frac{C_{mnl}c^3}{\hbar\mu_0}}    
\end{equation}

\begin{equation}
\label{ka}
   \kappa_a= \lvert\gagg\rvert B_0\sqrt{\frac{C_{mnl}\hbar c^3}{\mu_0}}    
\end{equation}

The parameters $\gagg,\rho_a, B_0, C_{mnl}$ are the axion-photon coupling strength, local axion density, external magnetic field, and a geometrical ``form factor" for the eigenmode, respectively.  The visibility function $\alpha(\omega)$, or the signal-to-noise ratio, is given by

\begin{multline}
\label{vis}
\alpha(\omega)=\lambda n_A\kappa_a\kappa_m\cdot\\
\Biggl\{(n_T+\frac{1}{2})\left[(1-\lambda)B(\omega)+\lambda\kappa_l\kappa_m\right]\\
+(n_b+\frac{1}{2})(1-\lambda+\frac{\lambda}{G_s})\lambda\cdot A(\omega)\Biggr\}^{-1},
\end{multline}

which can be used in $R=\int \alpha^2(\omega)d\omega$ to find the scan rate figure-of-merit:

\begin{widetext}

\begin{equation}
\label{rate1}
R_{amp}(\beta)=\frac{\pi n_A^2\kappa_a^2\kappa_l^{-1}}{2(n_T+\frac{1}{2})^2}\cdot\frac{\gamma^2\beta^2\sqrt{G_s}}{\sqrt{\lambda+G_s(\gamma+\lambda)(1-\lambda)/\lambda}\cdot\left\{\frac{(\beta-1)^2}{4}\left[\frac{\lambda}{G_s}+1-\lambda\right]+\gamma\beta+\gamma\frac{(1+\beta)^2}{4}\ \frac{1-\lambda}{\lambda}\right\}^{3/2}}
\end{equation}

\end{widetext}

where we further define %$\beta\equiv\kappa_m/\kappa_l$, 
$\gamma\equiv(n_T+\frac{1}{2})/(n_b+\frac{1}{2})$. This scan rate figure-of-merit will be normalized in later sections relative to the theoretical benchmarks.  Here we note that $R_{amp}$ scales as $|\gagg|^4$. 

\subsection{Linear Amplifier in a Standard Configuration}

We first consider the case in which the squeezer gain $G_s$ is unity and $\gamma=1$, setting the termination $T_b$ to be the same as the haloscope $T$. $R_{amp}$ then reduces to $\propto \beta^2/(\beta+1)^3$, which peaks at $\beta=2$ \cite{kenany17,malnou}. 
At large $\beta$, the scan rate scales as $\propto\kappa_l^{-1}/\beta\sim Q_0/(1+\beta)$.  To follow up a candidate signal, one seeks to maximize $\alpha(0)\propto \beta/(1+\beta)^2$ which peaks at $\beta=1$. All these are well-known results in the axion literature.

With $G_s=1$, one can either amplify-digitize both the $X$ and $Y$ quadrature or perform a single-quadrature ($X$) amplification.  The two-quadrature (phase-insensitive) amplification must add an additional half photon to the occupation number \cite{caves}, and the single-quadrature (phase-sensitive) amplification adds none.  However, as pointed out in \cite{malnou} this apparent advantage for single-quadrature noiseless amplification is canceled by $2\times$ fewer independent measurements and $2\times$ larger noise bandwidth created by the idler. Therefore, phase-sensitive amplification alone does not evade the standard quantum limit (SQL) at the system level. 

\subsection{Linear Amplifier in a Squeezed State Readout}
\label{sec:squeezed}

\begin{figure}
    \centering
    \includegraphics[width=0.8 \columnwidth]{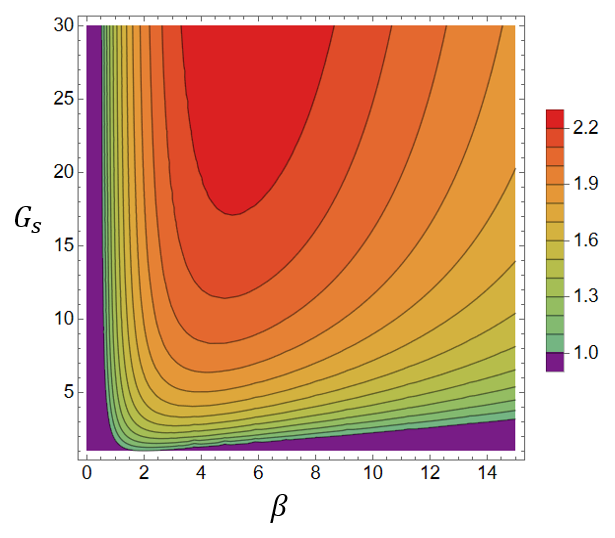}
    \caption{With $\gamma=1$ and $\lambda=\sqrt{0.69}$, the scan rate enhancement $E_t$ derived from Eqn(\ref{rate1}) reproduces results in \cite{malnou} for all $(\beta,G_s)$. The enhancement $E_t$ is independent of $n_T$ since both thermal and quantum noise can be squeezed.}
    \label{fig:etcheck}
\end{figure}

When the squeezer gain $G_s>1$, SQL can be evaded as pointed out in \cite{malnou} and realized in \cite{backes,jewell}. An enhanced scan rate is achieved by squeezing the photon noise from the termination ($n_b+1/2$) and increasing the coupling $\beta$.  As the cavity becomes overcoupled (larger $\beta$), it becomes more reflective.  As a result the amplifier sees more photon noise from the termination and less from the cavity. 
When $\gamma=1$, our Eqns (\ref{sigma})(\ref{vis}) recover Eqns A(23) and A(24) in \cite{malnou}.  When $\lambda=1$, Eqn \ref{rate1} reproduces their Eqn (2). Setting $\lambda=\sqrt{0.69}$, the scan rate enhancement $E_t$ given by the 2D function $E_t=R(\beta,G_s)/R(\beta=2,G_s=1)$ is plotted in Fig. \ref{fig:etcheck}, which accurately reproduces Fig. 2(b) in \cite{malnou}.  
%These give us confidence in the results when $\gamma \neq 1$.

While the above calculations are carried out using quantum mechanics, the final results can be understood classically aside from the $1/2$ photon corresponding to the vacuum state and the appearance of the Planck emission law. We point out the obvious fact that squeezing ($G_s>1$) improves the scan rate even when $n_T \gg 1$. This is quite clear since $n_T$ drops out completely in $E_t$ when the ratio is taken. The implication is that {\bf a squeezer improves the scan rate regardless of whether the experiment is cooled to the vacuum state}. In other words, ADMX at 1 GHz, or even DMRadio-GUT\cite{dmradiogut} at 1 MHz can still benefit from single-quadrature squeezing. While this broader applicability was already obvious from Eqn (2) of Malnou et al., it has not been widely appreciated by the axion community because of the emphasis on ``quantum'' state squeezing.  

%At critical coupling ($\beta=1$), the cavity is ``black" on resonance, 

The reduction of photon background cannot be done by attenuation, because any dissipation would induce emission (Kirchhoff's law), as is apparent from the effects of $\lambda$ in Eqn (\ref{vis}). 
Squeezing achieves this ``cooling" by de-amplifying the thermal (and vacuum) fluctuations in the quadrature being measured.  
Another way to achieve a similar effect is to simply cool the termination resistor to a lower temperature than the haloscope temperature ($T_b<T, \gamma>1$).  As discussed earlier, this situation happens because of cryogenic limitations such as the hot rod problem \cite{backes, jewell}.  In such scenario, $\gamma$ would be playing a similar role to $G_s$.

\begin{figure}
    \centering
    \includegraphics[width=0.8 \columnwidth]{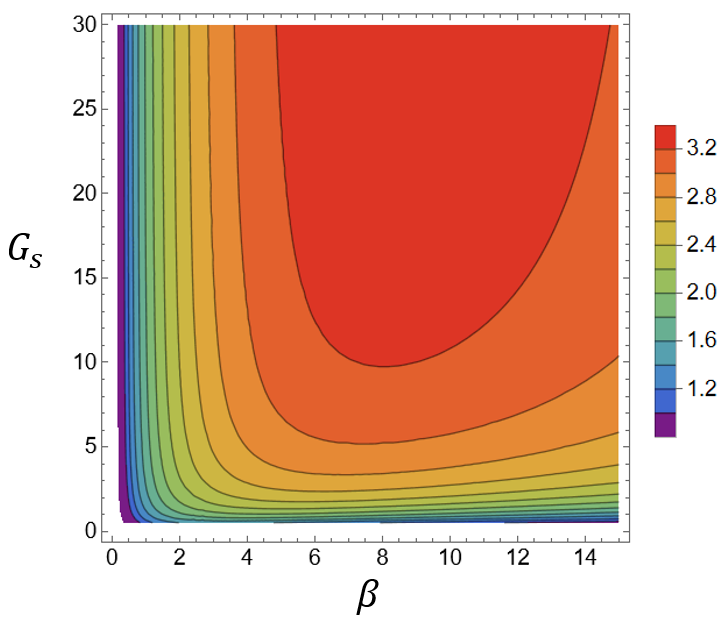}
    \caption{ Squeezing can be used in conjunction of having a lower $T_b$. 
 In this contour plot, we assume $\gamma=3.66$, baselining typical performance of ADMX. The efficiency $\lambda$ remains at $\sqrt{0.69}$. The scan rate peaks at $(\beta,G_s)\sim (8,20)$, giving an enhancement factor of 3.3$\times$.  ADMX typically operates at $\beta$=1.8 with no squeezing, $G_s=1$. }
    \label{fig:gamma3.66}
\end{figure}

The HAYSTAC cavity photons are at around 250 mK, contrasting the termination resistor's 61 mK \cite{backes,jewell}.  At 4.5 GHz, this corresponds to $\gamma=2.33$, already providing some scan rate enhancement even when there is no squeezing ($E_t\sim 1.55$ over $\gamma=1$). The ADMX\cite{admx21b} cavity photons typically sit at $\sim$300 mK, and the mixing chamber can be as cold as 80 mK. If the termination resistor can be cooled to 80 mK, it would correspond to $\gamma=3.66$ at 1 GHz. Assuming once again $\lambda=\sqrt{0.69}$, we obtain $E_t=R_{amp}(\beta=8,G_s=20)/R_{amp}(\beta=2,G_s=1)=3.3$, an enhancement in the scan rate by $3.3\times$ (Fig. \ref{fig:gamma3.66}).

The enhancement can be quite significant for MADMAX \cite{madmax_status_20}, which plans to run the haloscope at 4 K.  A compact dilution refrigerator could be used to cool only the termination connected through a circulator, providing a $\gamma$ factor significantly larger than 1.

\begin{figure}
    \centering
    \includegraphics[width=1 \columnwidth]{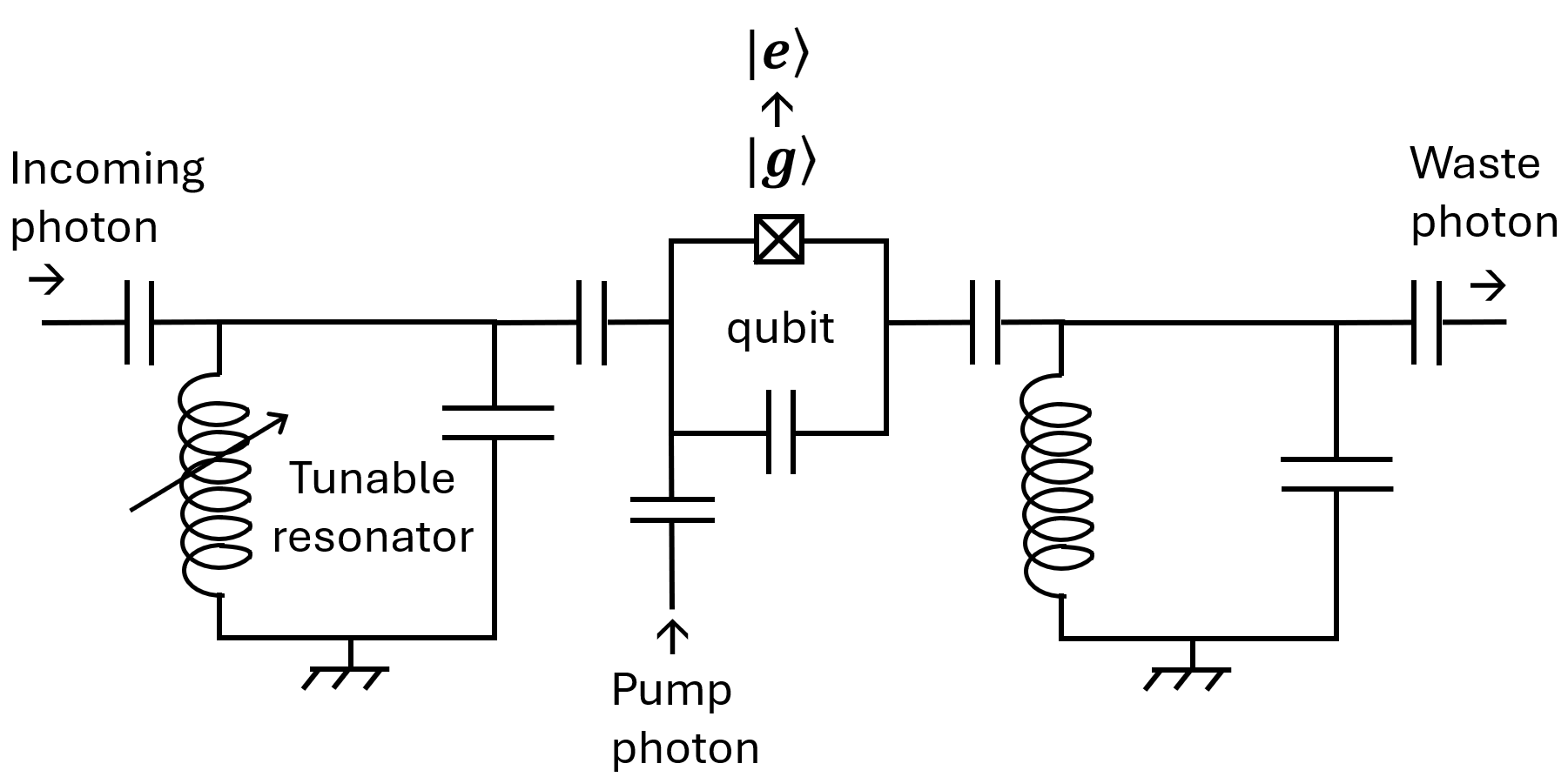}
    \caption{The SMPD \cite{braggio,lescanne,balembois,albertinale} consists of an intake resonator (``buffer") that defines the band of the incoming photons, a resonator that defines the outgoing mode (``waste"), and a transmon qubit coupled to both resonators. An incoming photon combines with a pump photon induce the qubit transition from $|g\rangle$ to $|e\rangle$, and the generation of a waste photon that subsequently dissipates .The schematic figure is adapted from \cite{albertinale}.}
    \label{fig:smpdsketch}
\end{figure}

\subsection{Microwave Photon Counting Detector}
\label{sec:smpd}

\begin{figure}
    \centering
    \includegraphics[width=0.6 \columnwidth]{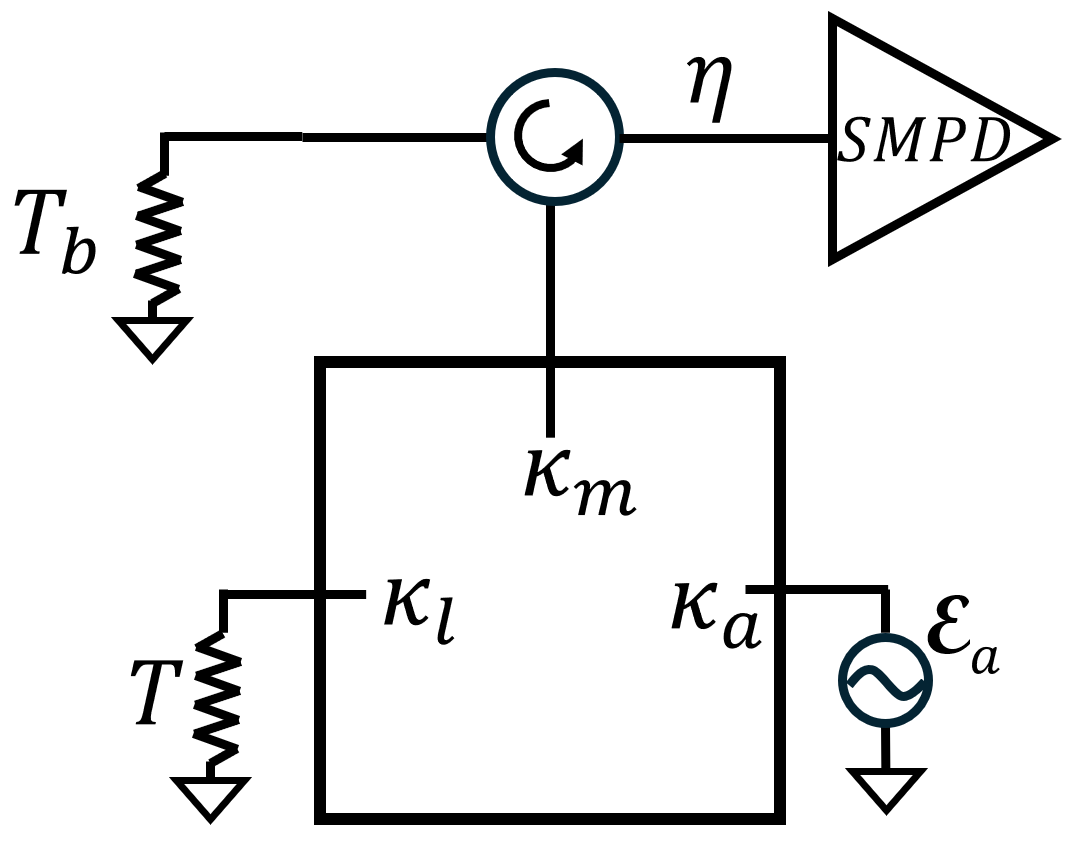}
    \caption{The experimental setup used in the analysis of \S 3C. In place of the amplifier, an SMPD (single microwave photon detector) is used for readout at the measurement port.}
    \label{fig:smpd}
\end{figure}

One can avoid the SQL by forgoing the phase information and by only {\em counting} the microwave photons \cite{lamoreaux}.  Notable examples of this approach are Rydberg atoms \cite{rydberg,rydberg2}, Penning traps \cite{penning}, and superconducting detectors \cite{schuster,dixit}.  In recent years, transmon qubits have emerged to be a leading technology to measure single photons in the microwave regime \cite{schuster,dixit,lescanne,albertinale,balembois}. In \cite{braggio}, such devices are coupled to a haloscope through a coaxial transmission line. This device architecture, which the authors called SMPD (single microwave photon detector), is particularly appealing because it creates physical separation between the superconductive sensors and the strong magnetic fields associated with an axion experiment.  The irreversible photon detection scheme also makes high-duty cycle integration possible.  A schematic of SMPD is shown in Fig. \ref{fig:smpdsketch}, along with a brief explanation of its operating principles. 

%\begin{widetext}
    
%\end{widetext}

For the purpose of this optimization/comparative study, we focus on microwave photon counters whose physical parameters are similar to the SMPDs reported in \cite{albertinale,balembois,braggio}. Although many of the results should be applicable to other types of microwave photon counters.  Reference \cite{rydberg2} considered several similar issues in the context of Rydberg-atom-based axion searches. We consider the system depicted in Fig. \ref{fig:smpd}, where the linear amplifier in Fig. \ref{fig:amp} is replaced by an SMPD.  The photon field detected is given by 

%\begin{widetext}

$$
\eta P(\nu)= \left\{n_T (1-\eta)+\frac{\eta}{B(\omega)}\left[n_T\kappa_l \kappa_m+n_b A(\omega)\right] \right\}\cdot h\nu 
$$

For single-moded optics, the noise equivalent power $\sigma$ associated with $P(\nu)$ consists of contributions from the Poisson noise and the Bose noise:

$$
\sigma^2 =\frac{1}{\tau}\left[
\int d\nu h\nu \eta P(\nu)+\int d\nu \eta^2 P^2(\nu) \right]
$$

Note that Eqns (9)(13) in \cite{lamoreaux}, Eqn (12) in \cite{richards}, and Eqn (41) in \cite{Zmuidzinas03} all contain the same expression.  The readers shall refer to these papers for the derivation.  The total noise background is given by 

%\begin{widetext}
    
\begin{multline}
\sigma_{tot}^2 =\frac{1}{\tau}\int_{\nu_c-\Delta \nu_{d}/2}^{\nu_c+\Delta \nu_{d}/2} d\nu h^2\nu^2 
\Biggl\{ n_T (1-\eta)\\
+\frac{\eta}{B(\omega)}\left[n_T\kappa_l \kappa_m+n_b A(\omega)\right]\\
+n^2_T (1-\eta)^2+\frac{\eta^2}{B^2(\omega)}\left[n^2_T\kappa^2_l \kappa^2_m+n^2_b A^2(\omega)\right]\\
+2n_T (1-\eta)\cdot\frac{\eta}{B(\omega)}\left[n_T\kappa_l \kappa_m+n_b A(\omega)\right]\\
+\frac{2\eta^2}{B^2(\omega)}\left[n_Tn_b\kappa_l \kappa_mA(\omega)\right]
\Biggr\}\\
\label{noise}
\end{multline}

%\end{widetext}

\begin{figure*}[t]
    \centering
    \includegraphics[width=1.7 \columnwidth]{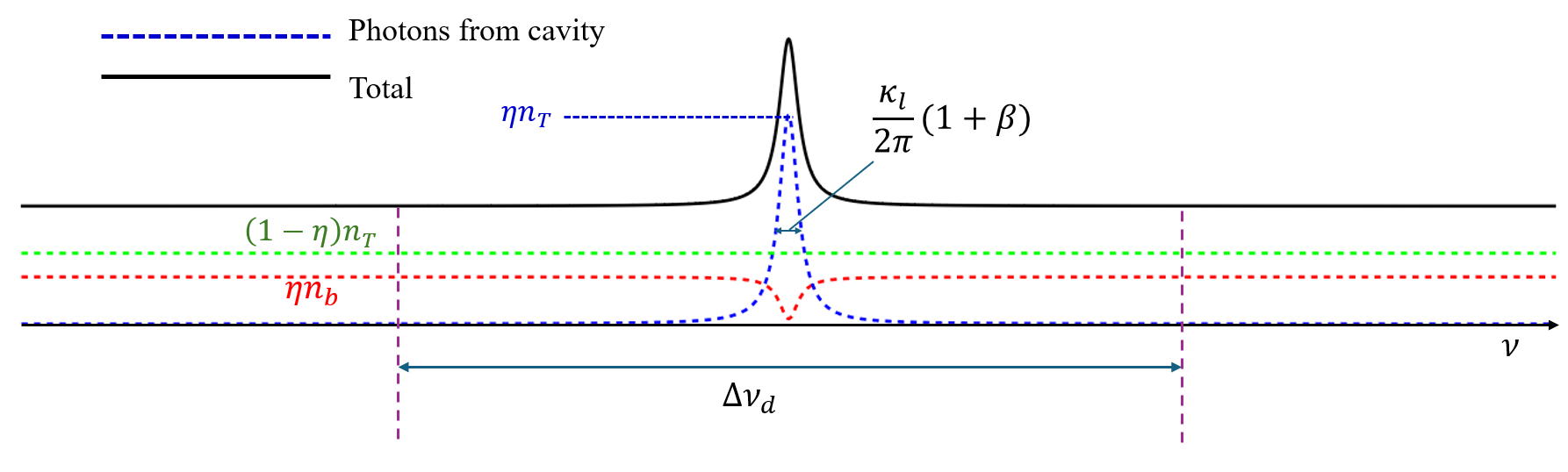}
    \caption{The contributions to photon noise near the resonance of the haloscope cavity.  For a photon counting detector, the noise within the bandwidth of the detector $\Delta \nu_{d}$ contribute to measurement uncertainties against a possible axion signal within the bandwidth of the haloscope, $\sim \kappa_l(2\pi)^{-1}(1+\beta)$. Based on the noise integral, Eqns (\ref{noise})(\ref{rate2}), the on-resonance cavity emission is independent of $\Delta \nu_{d}$, while the off-resonance contributions (emission from the termination and the lossy part of the transmission line) are proportional to $\Delta \nu_{d}$.  In many circumstances $\Delta \nu_{d}$ needs to be just large enough to cover the resonance and minimize the noise contribution from $n_b$ and $(1-\eta)n_T$. See text for a detailed discussion.}
    \label{fig:reflecivity}
\end{figure*}

The origin of various terms in the integral is clear when one analyzes first the Poisson ($\propto n$) portion of the $\sigma^2$ integral. With a photon counter that has a detection bandwidth $\Delta \nu_{d}$ around $\nu_c$, the integral takes place within $\nu_c\pm\Delta \nu_{d}/2$. It has a component that comes from the cavity emission, and an off-resonance component that comes from the ``background," including the coaxial termination and the lossy transmission line. See Fig. \ref{fig:reflecivity}.

Note that when the first stage readout is a linear amplifier with $G_a\gg 1$, the number of photons at detection is increased vastly and the Poisson noise becomes negligible (see \S 3C in \cite{Zmuidzinas03}). 
%The advantage of having no Poisson noise is wiped out at high frequency/low background ($n_T\lesssim 1$), when the presence of $1/2$ photon starts to decimate the linear amplifiers. 
Despite the advantage of eliminating the Poisson contribution, the increasing dominance of the $1/2$ photon term in the noise makes linear amplifiers highly undesirable toward high frequencies.  

We next include effects of a non-negligible dark count rate (DCR).  The dark counts in state-of-the-art SMPDs come from several sources \cite{lescanne,balembois}.  The first contribution to the DCR comes from the photon/microwave environment.  Usually this is characterized as a residual effective temperature in the photon field, denoted by $T_\gamma$ (with photon occupation number $n_\gamma$), in the 30--50 mK range. This photon noise would add to the existing background $n_b$ and $(1-\eta)n_T$.  The noise integral of all these background sources is proportional to  $\Delta\nu_{d}$. 

There are other sources of dark counts associated with the transmon qubit in the SMPD: (a) thermal excitation of the qubit, and (b) false positives in the determination of the quantum state, which is ultimately related to the qubit coherence time $T_1$.  This qubit-related component, which we call $\delta\nu_{DCR}$, is independent of $\Delta\nu_{d}$. The Poisson noise from $\delta\nu_{DCR}$ adds in quadrature to Eqn (\ref{noise}) as an additional variance outside the integral. 

Similarly to the noise, the average (axion) signal is given by $ n_A\kappa_a\kappa_m\eta\int_{-\infty}^\infty B^{-1}(\omega)(d\omega/2\pi) $. If $\Delta\nu_{d}$ is large enough to cover the Lorentzian line, we can use $\int_{-\infty}^\infty (\omega^2+a)^{-1}d\omega=\pi/\sqrt{a}$ and $\int_{-\infty}^\infty (\omega^2+a)^{-2} d\omega=\pi/(2a^{3/2})$ to obtain closed-form results. In this limit, we have 

$$\int d\omega A(\omega)/B(\omega)=2\pi\Delta\nu_{d}-\kappa_m\kappa_l\int d\omega /B(\omega),
$$
$$\int d\omega A(\omega)/B^2(\omega)=\int d\omega /B(\omega)-\kappa_m\kappa_l\int d\omega /B^2(\omega)
$$
\begin{multline*}
\int d\omega A^2(\omega)/B^2(\omega)=2\pi\Delta\nu_{d}-2\kappa_m\kappa_l\int d\omega /B(\omega)\\
+\kappa^2_m\kappa^2_l\int d\omega /B^2(\omega). 
\end{multline*}

Using these integrals recursively, we arrive at an expression for the frequency-integrated squared SNR $R_{SMPD}$, which is directly comparable to the scan rate figure-of-merit:

%\begin{widetext}
%$$
%\;\;\;\;\;\;\;\;\;\;\;\;\;\;\;\;\;\;\;\;\; \;\;\;\;\;\;\; 
%\;\;\;\;\;\;\;\;\;\;\;\;\;\;\;\;\;\;\;\;\; \;\;\;\;\;\;\; %\;\;\;\;\;\;\;\;\;\;\;\;\;\;\;\;\;\;\;\;\;\;\; 
%$$   
\begin{multline}
R_{SMPD}=2 n_A^2\kappa_a^2\kappa_l^{-1}\eta^2\beta^2(1+\beta)^{-2}\\
\Biggl\{\delta+D\Delta+E\beta(1+\beta)^{-1}
+F\beta^2(1+\beta)^{-3}\Biggr\}^{-1},
\label{rate2}
\end{multline}

%Note that $\Delta\nu_{d}/(\kappa_l\pi)$ is the bandwidth of the photon detector measured in terms of the unloaded linewidth of the cavity.

where the DCR appears as $\delta\equiv \delta\nu_{DCR}(\pi\kappa_l)^{-1}$ , a ``dark count rate" normalized by the unloaded cavity line width. 
$\Delta\equiv \Delta\nu_{d}(\pi\kappa_l)^{-1}$, is a similarly dimensionless measure of the detector bandwidth. The terms in the expression are organized based on their dependence on $\Delta$ and $\beta$.  The $D,E,F$ functions are:

\begin{multline*}
    D(\eta,n_T,n_b)=n_T(1\!-\!\eta)+n_b\eta\\
    +[n_T(1-\eta)+\eta n_b]^2+n_\gamma
\end{multline*}
\begin{multline*}
E(\eta,n_T,n_b)=2\eta\Biggl[n_T-n_b-2\eta n_b^2+2n_T^2(1-\eta)\\
+2n_Tn_b(2\eta-1)\Biggr]
\end{multline*}
\begin{equation*}
F(\eta,n_T,n_b)= 4 (n_T-n_b)^2\eta^2
\end{equation*}

%\color{red}
In the remainder of this section, various scaling relations are discussed. In \S III.B, the numerical values supported by the results achieved by ongoing experiments have been inserted into the equations to assess the experimental parameters required to reach various the main benchmarks.

Although the results seem verbose, the benefit of obtaining these analytic expressions becomes clear when various limits are taken.  
For example, whether $\delta\nu_{DCR}$ significantly affects the scan rate $R_{SMPD}$ depends on how it compares with $n_T\Delta\nu_{d}$. 
Completeness also helps clarify the dependence of the scan rate to the haloscope quality factor $Q$, a subject of great importance in the design of axion search experiments. 

Under truly background-free conditions, $n_\gamma, n_b,n_T\rightarrow 0$, or more precisely, $D\ll\delta/\Delta$, the axion experiment becomes limited only by the DCR. 

\begin{equation}
\label{case0}
    R_{SMPD} \rightarrow
    2 \pi n_A^2\kappa_a^2\eta^2\beta^2(1+\beta)^{-2}/\delta\nu_{DCR}.
\end{equation}

Intriguingly, the scan rate does not depend on $\kappa_l$ and increases weakly with $\beta$ in this limit.  We understand this observation as follows. Generally, a large $Q$ helps the scan rate by (1) resonantly enhancing the axion-to-photon conversion power and (2) reducing the bandwidth to background photons. On the other hand, high $Q$ hurts the scan rate by (3) reducing the search bandwidth and (4) restricting the output from the haloscope.  
%All of these effects have been fully incorporated into Eqn \ref{rate2}. 
With no background, (2) is gone, the effects of (1)(3)(4) cancel, and the benefit of high $Q$ disappears.  
%\color{red}Implications of this result to future axion searches are discussed in Section III.B.2. 
%\color{black}

\begin{figure*}[t]
    \centering
    \includegraphics[width=2 \columnwidth]{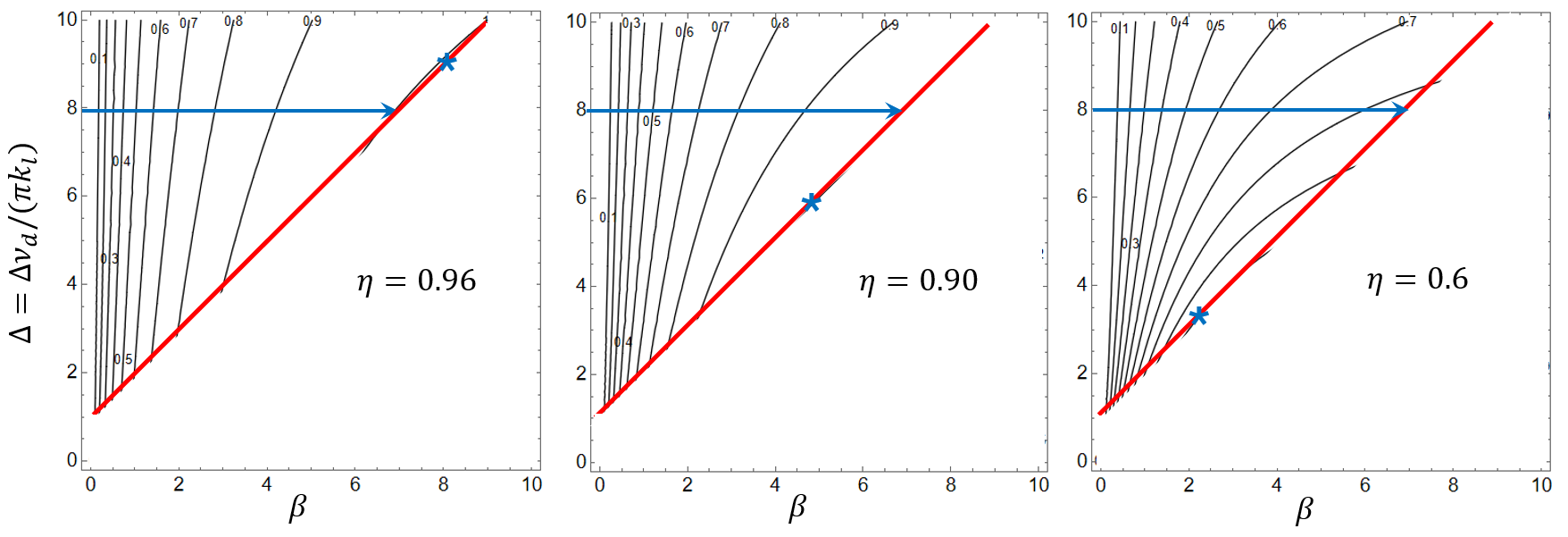}
    \caption{These contour plots show the scan rate $R_{SMPD}$ given in Eqn (\ref{rate2}) versus the coupling parameter $\beta$ and the photon counter bandwidth $\Delta=\Delta\nu_{d}(\kappa_l\pi)^{-1}$.  As discussed in the text, when $\Delta$ is fixed, the scan rate increases monotonically with $\beta$ as depicted by the arrow at $\Delta=8$ (or any other horizontal line). When $\Delta$ is engineered to be just large enough to cover the resonance, e.g., along the red line ($\Delta=(1+\beta)$), $R_{SMPD}$ maximizes at the locations of the asterisk:  $\beta\sim 8$ ({\em Left}) when the emission is dominated by the cavity (Eqn \ref{case1}), or $\beta\sim 2$ ({\em Right}) when the photon noise is dominated by the off-resonance background.  In between either limits, the scan rate maximizes at $\beta\sim 5$ ({\em Middle}).  The contours are normalized to the maximum in each plot. 
    The efficiency due to dissipation $\eta$ effectively changes the contribution from the off-resonance background photons. The SMPD ``dark count" due to the residual photons $T_\gamma$ has similar effects. The white-out regions in the lower right are where the SMPD bandwidth $\Delta\nu_{d}$ starts to be comparable to the linewidth $\nu_c/Q$. }
    \label{fig:betadelta}
\end{figure*}

When the noise is dominated by photons from the terminating load and other sources of off-resonance background (negligible $\delta\nu_{DCR}$ and $D\Delta\gg E\beta(1+\beta)^{-1}$), we have
%For $\eta=1$, the scan rate then reduces to :

\begin{equation}
\label{case2}
    R_{SMPD} \rightarrow
\frac{2\pi n_A^2\kappa_a^2\beta^2(1+\beta)^{-2}}
%{(n_\gamma+n_b+n_b^2)\Delta\nu_{d}}.
{D(\eta,n_T,n_b)\Delta\nu_{d}}.
\end{equation}

Although the scan rate also {\em appears} independent of $\kappa_l$ or $Q_0$, it is inversely proportional to the detector bandwidth. If the bandwidth can be engineered, it should be the narrowest possible to still cover the haloscope linewidth $\nu/Q_L$, i.e., $\Delta \nu_{d}\sim \nu\kappa_l(\beta+1)$. Equation (\ref{case2}) becomes proportional to $\beta^2(1+\beta)^{-3}$ in such global optimization, which peaks at $\beta=2$. In other words, the conventional wisdom for axion searches applies. %\color{red} 
Fig.~\ref{fig:betadelta} illustrates how this optimization takes place on the $\Delta-\beta$ plane under various experimental conditions.  In the SMPD scheme, $\Delta \nu_{d}$ is determined jointly by the linewidth of the buffer and the waste resonators (Eqn. D5 in \cite{balembois}).  

In this limit, a competitive axion search must involve synchronized tuning of the SMPD band and the haloscope resonance over a wide frequency range. So far, the state-of-the-art SMPD can only be tuned by $< 2\%$.  Significant development is still needed on this front.

%When $\eta<1$, the lossy part emits thermal radiation ($n_T$) with an emissivity $(1-\eta)$. Even if $n_b,n_\gamma \ll n_T$, $\Delta\nu_{d}(1-\eta)$ can become larger than $\kappa_l\eta$ and the off-resonance noise dominates the integral. The scan rate is again inversely proportional to the detector bandwidth $\Delta\nu_{d}$. The behavior of the $n_\gamma$ term is similar. 

Next we consider the cavity photon-dominated regime.  When $n_T\gg n_b \Delta$ and $\eta\sim 1$, the photon noise is dominated by the haloscope emission. Therefore,

\begin{equation}
\label{case1}
    R_{SMPD} \rightarrow
\frac{n_A^2\kappa_a^2\kappa_l^{-1}\beta^2(1+\beta)^{-2}}{n_T\beta(1+\beta)^{-1}+2n_T^2\beta^2(1+\beta)^{-3}}.
\end{equation}

In this limit, the bandwidth of the photon counter $\Delta\nu_{d}$ is not detrimental to $R_{SMPD}$ because the noise contribution is dominated by the Lorentzian integral around the line (Fig. \ref{fig:reflecivity}). 

%In the photon noise-dominated regime, it is easy to see that $\Delta \nu_{d}$ needs to be large enough to cover the potential axion signal in the resonance while minimizing the noise contribution from these background sources off-resonance.  

%Using numerical values from \cite{braggio}: at $\nu=7.37$ GHz, $Q_L=2.25\times 10^5$, $\Delta\nu_{d}=900$ kHz, efficiency due to loss $\eta=0.7$, we indeed find the photon counter's sensitivity to lossy emission dominates the noise contribution. 

A noteworthy result stemming from Eqns (\ref{rate2})(\ref{case0})(\ref{case2})(\ref{case1}) is their dependence on $\beta$. For photon counters with a given bandwidth $\Delta \nu_{d}$, the scan rate $R_{SMPD}$ increases monotonically in $\beta$ and asymptotically approaches a maximum, until the resonance width approaches $\Delta \nu_{d}$.  Although in practice, the rate increases minimally beyond $\beta>10$.
%When $\beta=9$, the rate is at 90\% of the maximum value. 
%To our knowledge this curious observation has not been reported in the literature. 

Consider the numerical values from \cite{braggio}: $\nu_c=7.37$ GHz, $Q_L=2.25\times 10^5$, $\beta=3$. The linewidth of the cavity is then 130 kHz, still substantially narrower than the SMPD bandwidth $\Delta\nu_{d}=$700 kHz (clear from Fig. 2a of \cite{braggio}). According to Eqn \ref{rate2}, by going to $\beta=10$ the scan rate can increase by 47\%. 

As mentioned earlier, if the SMPD bandwidth can be optimized together with the rest of the experiment, the answer might be different.  
%When the noise is dominated by the qubit-induced $\delta\nu_{DCR}$ or the haloscope emission line, one would still prefer the largest $\beta$ allowed by $\Delta\nu_{d}$. 
In Fig \ref{fig:betadelta}, we illustrate the optimization of $\beta$ in both the fixed-- and engineered-- $\Delta\nu_{d}$ scenarios under different loading conditions. 

%We plot the scan rate as a function of $\beta$ for several scenarios.  (either limits, different $n_T$, etc.)

%\begin{widetext}
    
%\end{widetext}
\section{Optimizing the Experimental Conditions}
\label{sec:compare}

\subsection{Scan Rate Comparison between AMP and SMPD}

We proceed to use Eqns (\ref{rate1})(\ref{rate2}) to compare the scan rates of two identical haloscopes read out by (1) a linear amplifier in a single-quadrature squeezed setup, (2) a photon counter such as the SMPD.

If the photon background vanishes, option (2) is vastly superior and the choice between the two is obvious. However, the main axion haloscope projects that had produced meaningful constraints on $\gagg$ still find the effective photon temperature in the cavities to be between 90 mK and 300 mK, elevated compared to the base temperature of the dilution refrigerator. 
At this temperature range the photon noise is not negligible.

\begin{figure*}
    \centering
    \includegraphics[width=2.0\columnwidth]{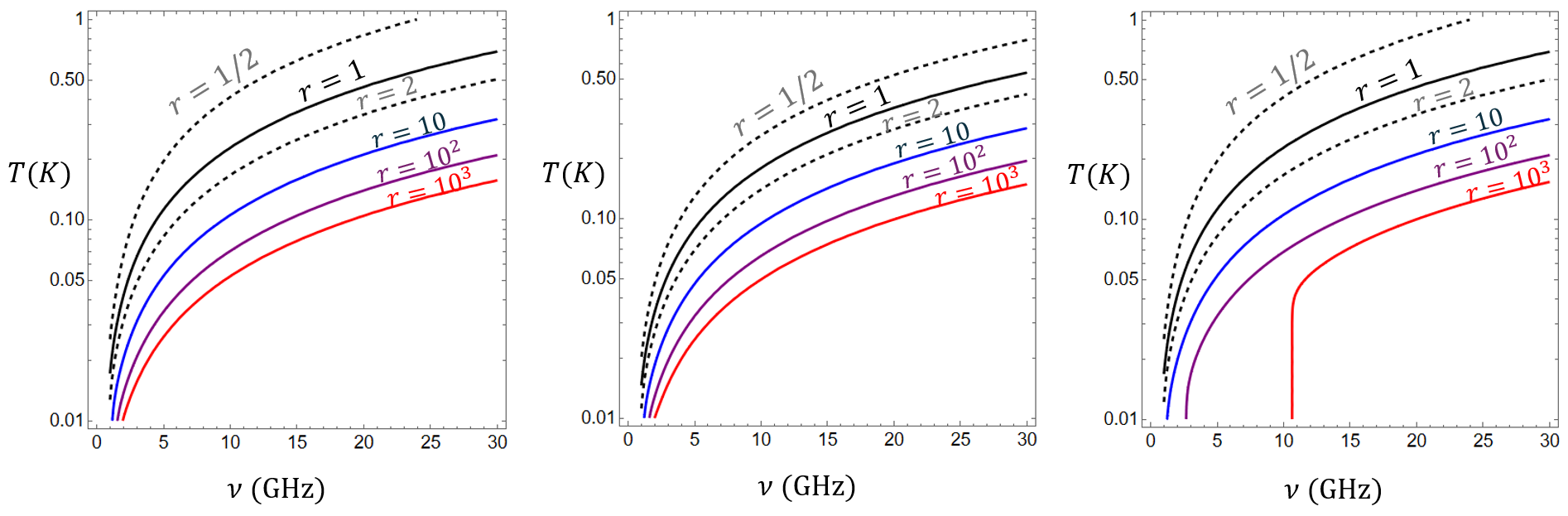}
    \caption{The contour plots for $r$, the scan rate ratio $R_{SMPD}/R_{AMP}$. {\em Left:} For a SQL amplifier operating in the standard configuration ($G_s=1,\beta=2$), $\delta\nu_{DCR}=100 s^{-1}$ for the SMPD.. {\em Center:} With an amplifier pair in a squeezed state configuration ($G_s=20,\beta=7$), $\delta\nu_{DCR}=100 s^{-1}$. {\em Right:} No squeezing on the amplifier but with a DCR of $\delta\nu_{DCR}=10^3 s^{-1}$.}
    \label{fig:compare}
\end{figure*}

To apply the results in the previous sections to a broader range of experiments, $T$ is set between 0.01 and 1 K. The termination temperature $T_b$ is assumed to be $T/3$ unless $T<$0.03, in which case $T_b$ bottoms out at 0.01 K. Using the haloscope temperature $T$ and the search frequency $\nu$ as parameters, we plot the contours of ratio $r\equiv R_{SMPD}/R_{AMP}$ (Fig. \ref{fig:compare}).  
In the left panel, the SMPD scan rate is compared against the baseline, no squeezing $G_s=1,\beta=2$. In the center panel, squeezing is turned on for the amplifier option. With the squeezing on, the contours shift to the right as expected. 

In the right panel, the SMPD has a dark count rate $\delta\nu_{DCR}=10^3s^{-1}$.  At the SQL, the equivalent DCR from an amplifier within the axion signal bandwidth is $\sim \nu/Q_a$, or about $10^4 s^{-1}$ at 10 GHz. In comparison, the DCR of the state-of-the-art SMPDs (few tens of $s^{-1}$) is already significantly better. That is why a DCR of $10^3 s^{-1}$ does not significantly change the overall conclusion except at the low frequency end. 

The $r=1$ contours represent the same scan rate for the amplifier and the SMPD. 
From these figures, it is clear that when $T<$ 150 mK and at high frequencies, SMPDs become overwhelmingly advantageous as $r$ becomes orders of magnitude greater than 1.  However, two things need to happen before the axion experiments broadly adopt SMPDs or similar microwave photon counters for the readout: (a) understanding the cause for the elevated photon temperature in the cavity and resolving the issue; and (b) upgrading the SMPDs to allow for robust, synchronized tuning with the haloscope.  Typically, an axion haloscope has a tuning range $\gtrsim$ 20\% and the microwave photon counter must match that.
Before these developments, one can continue to make good progress using SQL amplifiers.  Indeed, in Fig. \ref{fig:compare}, the $T=0.2$ K line crosses the equality curve ($r=1$) at 11 (9) GHz, and the $r=2$ line at 14 (12) GHz, with (without) squeezing.  

%review of the cryogenics situation for ADMX HAystac and 
%{\bf CAPP and QUAX }

%First we consider the impact on the scan rate from typical photon background originated in the haloscope and in the termination, as a function of frequency $\nu$, photon detector bandwidth $\Delta\nu_{d}$, haloscope temperature $T$, dissipative loss ($1-\eta$), and the termination temperature $T_b$. 

%State-of-the-art SMPDs have already achieved a DCR $\lesssim 100\;s^{-1}$.  At 20 mK the photon background rate $n_T\Delta\nu_{d}\ll 1\; s^{-1}$.  Therefore, DCR would be the limiting factor at this temperature at around 7 GHz \cite{braggio}.  However, at 50 mK, $n_T\Delta\nu_{d}$ already exceeds 800 $s^{-1}$. 
%Therefore, the DCR should not the limiting factor for current science grade experiments with mechanical tuning elements in which the effective cavity photon temperature $T$ is around 250 mK and the base temperature is $\gtrsim$ 50 mK.

%\end{widetext}

%\section{Discussion}

\begin{table}%The best place to locate the table environment is directly after its first reference in text
\caption{\label{tab:table1}%
The parameters used in Fig. 8--11. The scan rates have been normalized by achieved $\gagg$ limits in \cite{backes}.  Therefore, these contour plots have built in realistic integration and scanning inefficiencies -- including calibration, tuning, thermal settling, etc. 
%(\ref{fig:compare})(\ref{fig:compare1})(\ref{fig:compare2})(\ref{fig:compare3})
}
\begin{ruledtabular}
\begin{tabular}{llcr}
Parameter& 
Description& 
Value &
Unit 
\\
%\multicolumn{1}{c}{\textrm{Decimal}}&
%\textrm{Right}\\
\colrule
$Q_0$ & quality factor \cite{backes} &$4.7\!\cdot\! 10^4(\nu/4.14 {\rm GHz})^{-2/3}$&-\\
$V_{h}$&cav. volume \cite{backes}& 3.94$(\nu/4.14{\rm GHz})^{-3}$&$l$\\
$V_{vera1}$&vol. VERA-1& $36(\nu/4.14{\rm GHz})^{-1}$& $l$\\
$V_{vera2}$&vol. VERA-2& $27(\nu/4.14{\rm GHz})^{-0.5}$& $l$\\
$B_0$&magnetic field&8&T\\
$T_b$ & if $T>0.03$ K & $T/3$ &  \\
$T_b$ & if $T<0.03$ K & 0.01 & K\\
$\eta$& efficiency& $\sqrt{0.7}$ &-\\
$\Delta\nu_d$& detector BW & $7\cdot 10^5$ & Hz\\
$\Delta\nu_d$& ibid., for Fig. \ref{fig:compare3} & $\nu/5$ & -\\
\end{tabular}
\end{ruledtabular}
\end{table}

\subsection{Reaching the KSVZ/DFSZ Limits for the Post-Inflationary Frequencies}\label{sec:dfsz}

The original Strong CP-resolving axion model was quickly ruled out by data. The minimal extension that is still consistent with the data is known as the KSVZ model after its proposers Kim \cite{kim}, Shifman, Vainshtein, and Zakharov \cite{shifman}. 
Another benchmark model that had received broad recognition is the DFSZ model proposed by Dine, Fischler, Srednicki \cite{DINE1981199}, and Zhitnitsky \cite{Zhitnitsky:1980tq}.  The DFSZ model predicts a coupling strength ($\gagg$) 2.6 $\times$ smaller than the KSVZ value \cite{marsh}.  
%Both models predict $\gagg$ to scale as $\propto \nu_a$. 

So far the ADMX and CAPP experiments have reached the DFSZ theoretical benchmark at around 1 GHz. The only other project nearing the theoretical limits is HAYSTAC \cite{backes,jewell}, coming within a factor of 2 of KSVZ over $\sim$ 2--3 \% of bandwidth between 4--5 GHz.  

In this study we are most interested in the frequency range $\nu_a$ between 1--30 GHz (cm-wave), corresponding to the post-inflationary axion production scenario 
\cite{marsh,Borsanyi2016,Klaer_2017,Graham_2018,Takahashi_2018,Buschmann2022}.
One main difficulty in scaling the $\sim$1 GHz experiments to to high frequencies is due to loss of volume. From Eqns (\ref{na})(\ref{rate1}), the scan rate is $\propto V^2$. 
Naive dimensional scaling of the cavity geometry in all three dimensions leads to a steep scan rate reduction at high frequencies ($\propto \nu^{-6}$).  In addition, $Q_0$ (or equivalently $\nu\kappa_l^{-1}$) decreases at high frequency due to the anomalous skin effect as $\sim \nu^{-2/3}$.  

As discussed in \S\ref{sec:squeezed}, operating the receiver in a squeezed state can improve the scan rate beyond the SQL. However, the main limitation of this approach comes from the dissipative loss between the squeezer and the haloscope, in particular from the circulator. In practical systems the gain in scan rate is often limited to around a factor of 2.  While this is significant enough to implement on working projects, the sensitivity gap to DFSZ at $>$ 5 GHz is too vast for this method to overcome.  In principle, the gain could be unlimited if the circulator is lossless, or can be avoided altogether.  R\&\!D efforts are being made to pursue this but should be characterized as in early stages. 

It is possible to improve the scan rates by significantly increasing the quality factor $Q$ using superconductive coatings, such as niobium alloys \cite{cervantes} or tapes of high $T_c$ superconductors. The most impressive results so far are those by the CAPP group using high $T_c$ superconducting tapes that has been shown to survive in strong $B$ fields.  It is unclear whether superconducting high-$Q$ cavities can effectively work with photon counting detectors.  As discussed in \S\ref{sec:smpd}, the frequency and the bandwidth of SMPD should match those of the haloscope in the off-resonance background-dominated regime. Synchronized tuning over a wide frequency range will be extremely challenging for these ultra high-$Q$ systems.  In the DCR-dominated regime, the scan rate becomes independent of $Q$ (Eqn. \ref{case0}), rendering null the benefits of having high quality factors.  Even restricting to $<$ 5 GHz and operating with an amplifier face significant challenges in microphonics and cryogenics when it comes to tuning a system with $Q_L\gtrsim 10^6$ over an appreciable search range. 

Other groups have proposed to use extremely high-field magnets for axion searches, motivated by the strong dependence of the scan rate to $B_0$ i.e., $R\propto B_0^4$. Advanced magnet engineering is used to produce a field strength 20 T or higher, leading to a seemingly vast advantage over standard NbTi magnets at around 8--9 T.  One challenge with this approach is that the very high-field magnets use resistive coils which incur staggering power consumption.  Reference \cite{grahal} reports the 43 T-field requires power consumption exceeding 20 MW, which seems prohibitive for typical scales of axion experiments.  The CAPP group reports continuous operation of a lower power 18T HTC magnet \cite{capp18T}.  Nevertheless, these high-field magnets often come with relatively small bore diameters, resulting in a figure-of-merit $B_0^2V<0.3$ T$^2m^3$, substantially lower than that of a 8--9 T magnet with an 800 mm-bore.  
%However, if the efforts to develop high-volume ($V\gg \lambda^3$) haloscopes are unsuccessful, going to high-field and small-bore should be revisited.

Based on these considerations, we focus the subsequent discussions on two R\&\!D thrusts aiming to reach DFSZ for much of the post-inflationary mass range: (1) breaking the volume scaling of $\nu^{-3}$ and vacuum state squeezing; (2) developing SMPD for axion dark matter searches. In the next two sections we discuss these approaches in turn. 

\begin{figure*}
\centering 
\includegraphics[width=6.6in]{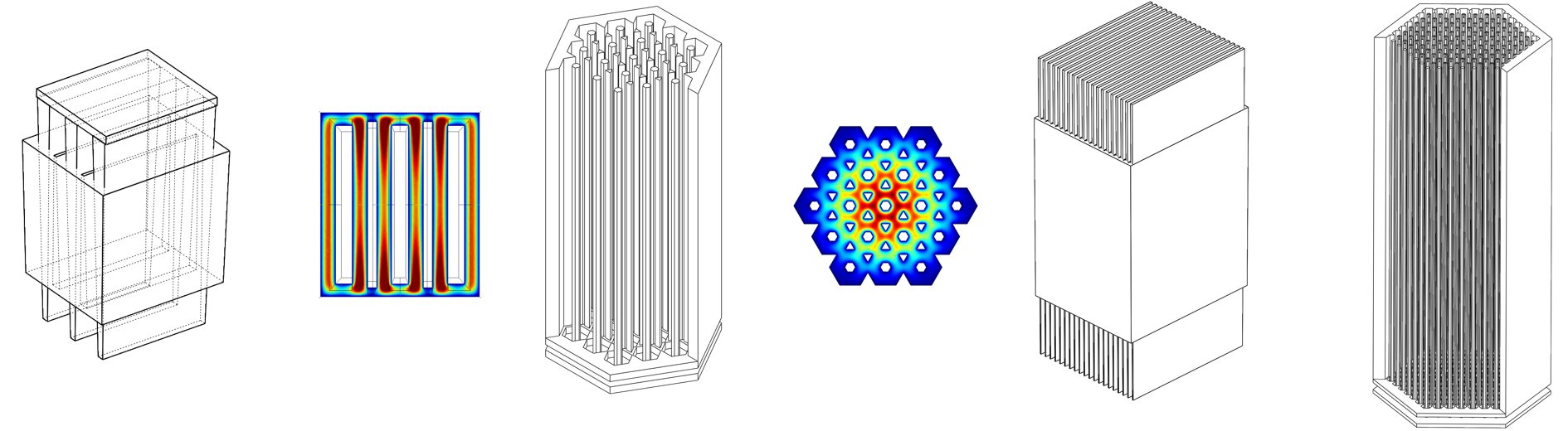}
\caption{\label{fig:vera} 
The VERA haloscopes. From left to right: a triple-wedge cavity (5--7 GHz), the cross section view of its field distribution \cite{Kuo_2020,Kuo_2021,dyson}; a beehive cavity consisting of overlapping cylinders (5--7 GHz) \cite{withers}, its field distribution; a notional 20-wedge cavity at $\sim$ 20 GHz, and a beehive cavity at $\sim$ 20 GHz. 
}
\end{figure*}

\subsubsection{Volume Enhancement and Squeezed Amplification}

The loss in volume at short wavelengths ($\lambda$) can be circumvented by adopting novel resonators that fill a volume $\gg\lambda^3$.  In \cite{Kuo_2020,Kuo_2021,withers}, we propose two such designs: the thin-shell ``wedge" cavity and multi-cell ``beehive" cavity (Fig.~\ref{fig:vera}).  These geometries can scale up in volume at a fixed resonant frequency and have $Q$'s similar to conventional cylindrical cavities. The wedge design has been experimentally demonstrated at 7.5 GHz \cite{dyson}. This approach is being vigorously pursued by the the ADMX-VERA (Volume Enhanced Resonant Axion) working group.  Haloscopes based on non-cavity resonators can also achieve similar volume enhancements. The most well developed concepts are the plasma haloscopes \cite{plasma,plasma3} and the dielectric disk haloscopes \cite{madmax_17,madmax_status_20}.

For the purpose of readout comparison, we use the VERA concept as the baseline. Much of the results and conclusions are applicable to other volume-enhanced haloscopes.  Scaling from \cite{backes}, we assume VERA can increase the volume of the resonant cavity at 4.14 GHz by a factor of 7--8, {\em and} change the frequency scaling from $V\propto \nu^{-3}$ to $V\propto \nu^{-\alpha}$, with an $\alpha<$ 3.  In principle, an ideal volume-filling multi-wedge cavity or a beehive cavity can completely eliminate the frequency dependence, corresponding to $\alpha=0$. However, there are still uncertainties regarding how this technology scales in frequency, despite great promises shown at $<$ 8 GHz. Practical limitations at shorter wavelength might induce phase errors that lead to lower form factors, and effectively, loss in volume. Recognizing that, we consider two stages in the VERA program: VERA-1 with $\alpha=1$ and VERA-2 with $\alpha=0.5$. Both are normalized at 7 GHz using a 3-wedge thin-shell cavity as the baseline model \cite{dyson}.  These two scenarios correspond to retaining $\sim$20\% and $\sim$50\% of the effective volume at 30 GHz compared to 7 GHz. 

%(6.1 l at 6.5)
%talk about vera-1 and vera-2. 

To make reliable forecasts, we normalize the scan rate expressions obtained in earlier sections by {\em achieved} exclusion limits reported by HAYSTAC in \cite{backes}.  In 105 days of operation, they ruled out $\gagg$ above 1.38 KSVZ at 90\% confidence level over 1.76\% of bandwidth around 4.14 GHz. The receiver is operated in a squeezed state setup with $G_s=21$ and $\beta=7.1$. The cavity photon temperature $T$ is measured to be 250 mK and the termination $T_b$ is at 61 mK. These parameters are used in normalizing the amplifier scan rate $R_{AMP}$ using Eqn \ref{rate1}, and by extension, Eqn. \ref{rate2}.  

For convenience, {\bf we normalize the ``KSVZ", ``DFSZ" and other benchmark contours to correspond to 100 days of operation over a 5\% fractional bandwidth.}  The HAYSTAC results have been obtained after several years of optimization.  What they have achieved should be considered results of a realistic experimental duty cycle, and actual sensitivity and efficiencies of a well-tuned axion experiment using modern cryogenic and readout systems.  These reality-calibrated rates should be a reliable forecast for reaching particular benchmarks over a bandwidth of 30\% after a few years of operation. 

\begin{figure*}
    \centering
    \includegraphics[width=2\columnwidth]{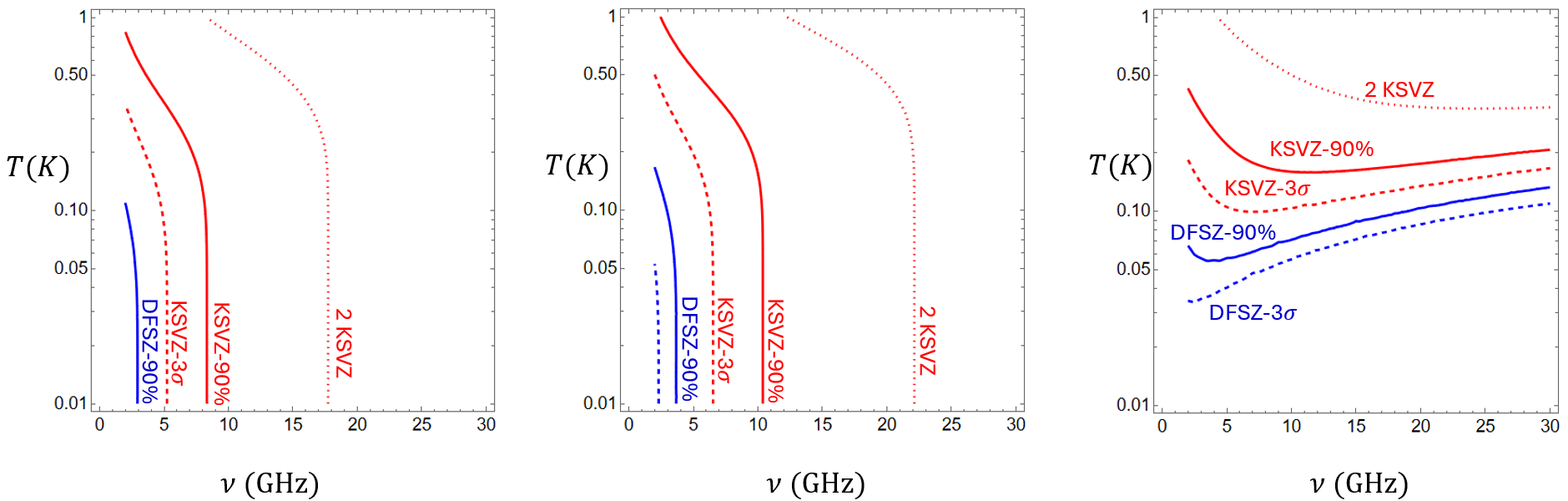}
    \caption{Combining the VERA approach and quantum techniques, it is possible to cover most of the post-inflationary axion mass range ($>$ 30 GHz) to DFSZ {\em Left:} VERA-2 cavity with SQL readout. The contours correspond to scan rates to reach DFSZ at 3$\sigma$ (blue dashed), DFSZ at 90\% confidence level (blue), KSVZ at 3$\sigma$ (red dashed), KSVZ at 90\% (red), 2$\times$KSVZ at 90\% (red dotted), and completing a 5\% search bandwidth in 100 days of operation around that given frequency. {\em Center:} VERA-2 cavity with the squeezed state readout ($G_s=20,\beta=7$). {\em Right:} VERA-2 cavity with SMPD, DCR=100 $s^{-1}$. {\bf Other experimental parameters are listed in Table I}.}
    \label{fig:compare2}
\end{figure*}

%\color{red}
Using the assumptions in Table I., it can be shown that either VERA-1 or VERA-2 can reach the KSVZ benchmark to 8 GHz with SQL amplifiers.  Additionally, 0.5 KSVZ is reachable by VERA-1 up to 15 GHz and VERA-2 up to 18 GHz (Fig. \ref{fig:compare2}, left panel).  The presence of $1/2$ photons in the noise creates two limiting effects that are noticeable: the inability to reach DFSZ above $\sim$ 4 GHz even with VERA-2, and the ineffectiveness in improving the experimental reach by cooling the cavity below 100 mK. 

In the middle panel of Fig. \ref{fig:compare2}, squeezing is enabled with VERA-2 with the standard parameters $G_s=20,\beta=7$. It produces the expected results of shifting the contours toward high frequencies. DFSZ is reachable in this case up to 4 GHz, while KSVZ can be reached $<$ 10 GHz.

\begin{comment}
    
\begin{figure*}
    \centering
    \includegraphics[width=2\columnwidth]{T_vs_nu_amp_haystac_vera_G1_beta2_3sigma.png}
    \caption{This set of contour plots illustrate how high-volume cavities proposed in \cite{Kuo_2020,Kuo_2021,dyson,withers} can greatly improve the scan rate of a otherwise standard SQL haloscope experiment. The contours correspond to scan rates to reach DFSZ at 3$\sigma$ (blue dashed), DFSZ at 90\% confidence level (blue), KSVZ at 3$\sigma$ (red dashed), KSVZ at 90\% (red), 2$\times$KSVZ at 90\% (red dotted), and completing a 5\% search bandwidth in 100 days of operation around that given frequency.   ({\em Left}) With the cavity volume $V_h$ \cite{backes}; ({\em Center}) with a high-volume cavity demonstrated in \cite{dyson} and a scaling of $V\propto\nu^{-1}$ (VERA-1); ({\em Right}) with a volume scaling of $\nu^{-0.5}$ (VERA-2). {\bf Other experimental parameters are listed in Table I}. Such a program is a robust proposal to reach KSVZ up to about 8 GHz with {\em existing} sensor technology. }
    \label{fig:compare1}
\end{figure*}

\end{comment}

\begin{figure*}
    \centering
    \includegraphics[width=2\columnwidth]{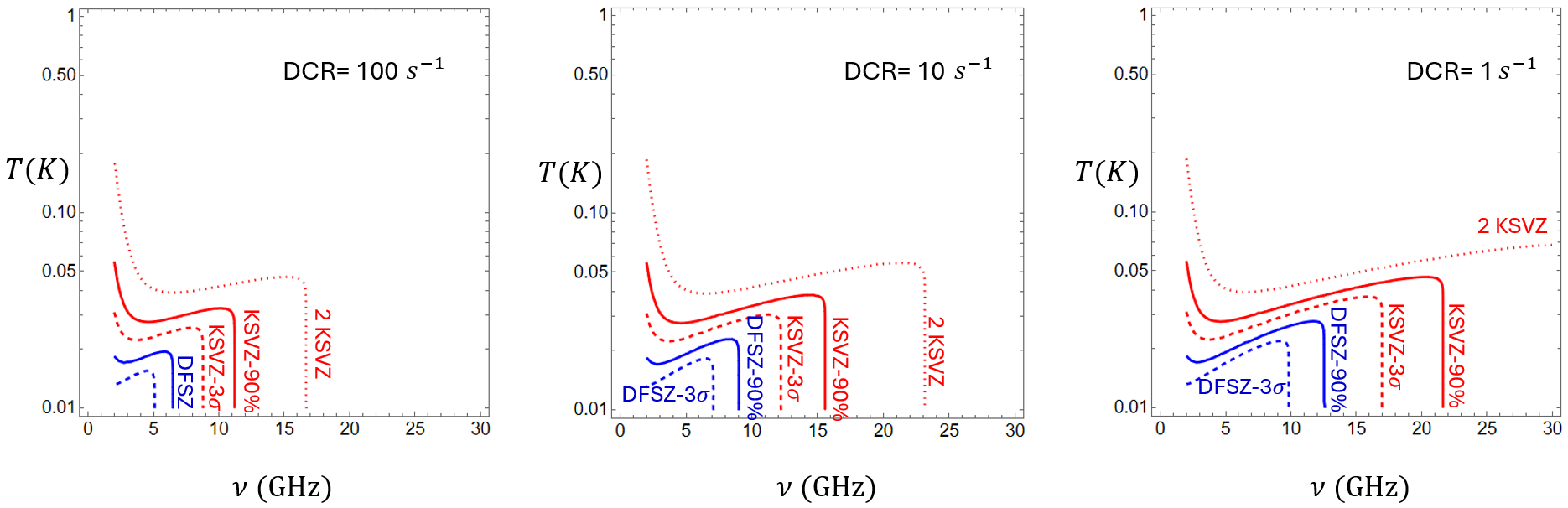}
    \caption{It is possible to significantly improve the scan rate to reach DFSZ up to 12.5 GHz even if $V\propto \nu^{-3}$. This can be done by (1) reducing the haloscope photon temperature to $\lesssim$ 30 mK to be in the background-free regime, and (2) progressively reducing the dark count rate of the SMPD. From left to right shows the experimental reach using a series of cavities directly scaled up in frequency from \cite{backes}, with DCR from $10^2$, 10, to 1 $s^{-1}$. In these plots, broadband operation of SMPD is assumed ($\Delta\nu_d/\nu=0.2$). In each line, the turning point from horizontal to vertical branches marks the $\gagg$ reach in the background-free regime. These points collectively trace out the $T=h\nu/k$ curve. }
    \label{fig:compare3}
\end{figure*}

\begin{comment}
\begin{figure*}
    \centering
    \includegraphics[width=2\columnwidth]{T_vs_nu_scaledhaystac_dnu_3sigma.png}
    \caption{This set of figures demonstrate the effects of lowering the detector bandwidth $\Delta\nu_d$ with DCR fixed at 1 $s^{-1}$. From left to right, $\Delta\nu_d/\nu=0.2$, $10^{-4}$, $10^{-5}$. In the right panel the quality factor is increased from the default (Table I) by a factor of 20. The performance of the experiment remains unchanged in the background-free regime when $T\lesssim$ 30 mK. Higher $Q$ and narrower $\Delta\nu_d$ improve the scan rate in the photon noise-dominated regime: $\nu<$ 5 GHz , $T>$ 50 mK.}
    \label{fig:compare4}
\end{figure*}
\end{comment}

\vspace{-0.5 cm}
\subsubsection{Using SMPD in Axion Searches}

Above 4 GHz, DFSZ is stubbornly out of reach even for VERA-2 {\em with} vacuum state squeezing.  The real game changer at these high frequencies must be the introduction of SMPD in axion searches.  In the right panel of Fig. \ref{fig:compare2} we show the combination of VERA and SMPD, which shows DFSZ becoming accessible for the entire $<$ 30 GHz range. Impressively, this is achieved even with a 100 $s^{-1}$ DCR in the detector and the cavity running at $\gtrsim$ 100 mK.

%\color{red}

In \S\ref{sec:smpd}, we observe three noise regimes for SMPD. 
In the off-resonance background dominated regime (such as in Fig. \ref{fig:betadelta} right panel), the scan rate is optimized with high $Q$, $\beta=2$ and the SMPD bandwidth $\Delta\nu_d$ just large enough to cover $\nu_c/Q_L$. In the on-resonance cavity emission dominated regime (Fig. \ref{fig:betadelta} left panel), the scan rate increases with $\beta$ but reaches a diminishing return when $\beta\sim 10$. 
In both cases, the SMPD band must be tuned together with the haloscope resonator, which usually has a frequency range $\gtrsim$ 20\%.  In a science integration, the operating efficiency would severely suffer if the tuning range of the detectors is too small compared to the haloscope.  We consider a 20\% frequency tunability a necessary requirement for adopting SMPDs in axion searching programs aiming at reaching the theoretical benchmark over significant mass ranges. 
As Fig. \ref{fig:smpdsketch} depicts, the input band is determined by the tunable resonator on the left, dubbed the ``buffer" resonator in \cite{lescanne,balembois,albertinale}. In the current implementation, the tuning range is still limited to $<3\%$. 

The third regime corresponds to the background-free limit, where 
%all thermal backgrounds -- $n_T,n_b,n_\gamma$ are controlled to be negligible compared to the equivalent temperature of the DCR $\delta/\Delta$, 
Eqn \ref{case0} applies. In this case, the scan rate of the haloscope is independent of $Q$ in the search mode.  Therefore, the SMPD band just needs to cover the cavity haloscope's frequency range, e.g., 20\%, and does not need to be tunable.  This is also the limit in which the BREAD experiment aims to operate, using ultra-low background Cooper pair-breaking detectors of single infrared photons \cite{bread}. Note that a higher $Q$ offers a measurement of the axion mass (to a precision of $1/Q$) in the event that a positive detection is made.

\begin{figure*}
    \centering
    \includegraphics[width=1.4 \columnwidth]{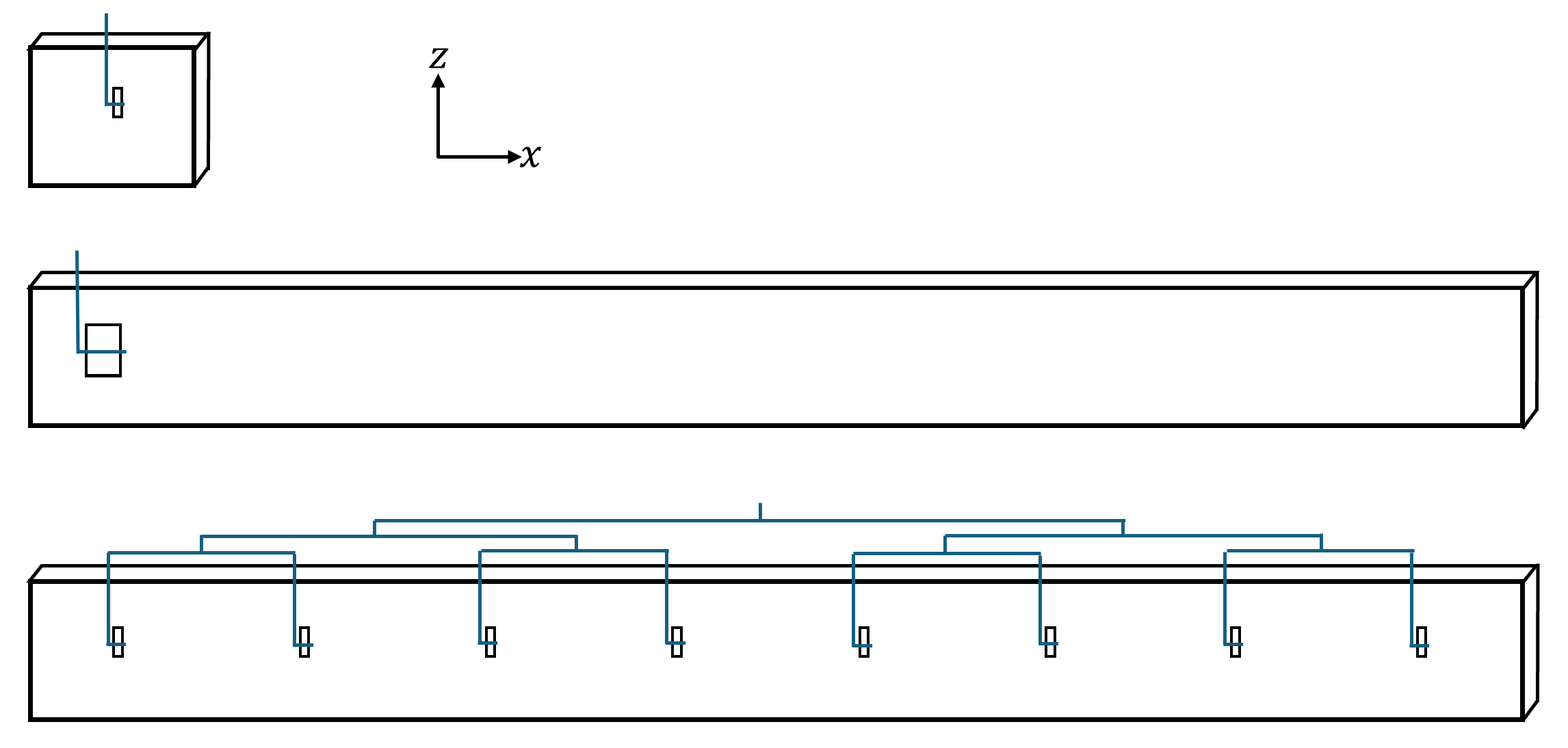}
    \caption{({\em Top}) A rectangular cavity coupled to a read out port (or a slot antenna) with $V\sim\lambda^3$. We are interested in reading out the fundamental mode that is polarized in the $x$ direction whose frequency is set mostly by the smallest dimension ($\lambda/2$). We assume the port here provides the optimal $\beta$ parameter.  ({\em Middle}) The volume of the cavity is grown ($V\gg \lambda^3$) by increasing the its dimension in $x$ \cite{Kuo_2020}.  The fundamental mode with uniform polarization in  $x$  still exists, however to maintain the same $\beta$, one must increase the coupling strength of the port proportionally. Doing so with a single large port would drive the mode away from the port instead of achieving the goal (Fig. 13). ({\em Bottom}) Instead, one must implement an array of ports distributed throughout the volume of the cavity and coherently add up the signal in a phase-matching summing network.  See text for more discussions.}
    \label{fig:couplingvera}
\end{figure*}

In Fig. \ref{fig:compare3}, we see the SMPD-based axion experiment enters the background-free regime when $T$ is less than about 30 mK, depending on the frequency.  The SMPD band is fixed, with a bandwidth of 20\%. The haloscope cavity is tunable within this band, with $\beta=10$.  After completing each 20\% range, a new set of cavity and SMPD are swapped in.  The three panels represent DCR at 100, 10, and 1 $s^{-1}$. The DFSZ is reachable up to 12 GHz at $\delta\nu_{DCR}=1 \;s^{-1}$.  This set of forecasts assume the haloscope volume scales as $\nu^{-3}$ from \cite{backes}, .i.e., without using any of the volume enhancement resonators discussed in the previous section.  This device-heavy approach carries its own set of risks and benefit compared to VERA and should be viewed as highly complementary. 

\begin{comment}
Once in the background-free regime, the scan rate becomes independent of the detector bandwidth and the haloscope quality factor, as illustrated in Fig. \ref{fig:compare4}. From left to right, the SMPD bandwidth is changed from 20\% to $10^{-4}$, and $10^{-5}$, all with the dark count rate of 1 $s^{-1}$. In the right panel, the quality factor is increased from the default (Table I) by a factor of 20 and $\beta$ reduced from 10 to 2. The main effect of high $Q$ and narrower SMPD bandwidth is the improved scan rate at lower frequencies ($<$ 5 GHz) and higher system temperature ($>$ 50 mK). The performance of the experiment remains unchanged when $T\lesssim$ 30 mK. 
\end{comment}

\subsection{Achieving Strong Coupling in High-Volume Axion Haloscopes}

 In the first experimental demonstration of VERA, the coupling parameter $\beta$ with a single antenna was measured to be $\sim 0.1$ \cite{dyson}, substantially lower than the optimal value of 2.  In earlier sections of this paper, we discuss many scenarios in which overcoupling beyond $\beta=2$ is desired. It is worth investigating how one can achieve sufficient coupling in a large-volume cavity.

%\color{red}
\begin{figure*}
    \centering
    \includegraphics[width=0.8\linewidth]{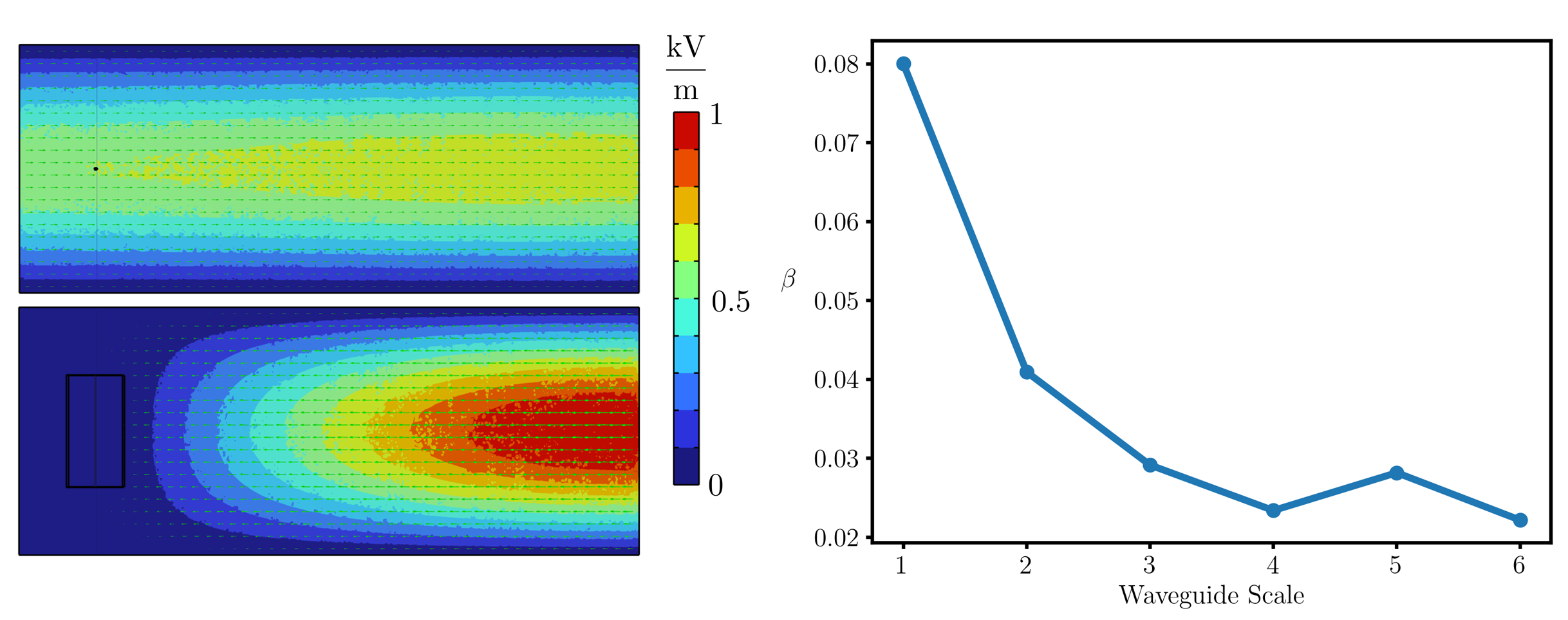}
    \caption{(\textit{Left}) Electric field contour plots simulated in COMSOL-RF for a $V \gg \lambda^3$ rectangular cavity with two different readout port sizes. When a small readout port is added (top contour plot), coupling is low because energy losses at the cavity walls are proportionally larger than the energy extracted by the port. The cavity mode also begins to localize away from the port location. When the port is grown in an attempt to increase the extracted energy (bottom contour plot), this mode localization increases, leading to further degradation of the coupling. For both plots, the cavity dimensions were $1000~\textrm{mm}~\times~400~\textrm{mm}~\times~15~\textrm{mm}$, and the resonant frequency of the mode was 10 GHz. The readout ports were simulated using rectangular waveguides.  (\textit{Right}) Receiver chain coupling $\beta$ achieved with different sized readout ports for the same $V \gg \lambda^3$ cavity. The scale 1 waveguide had a broad wall length of 30 mm for a cutoff frequency of 5 GHz. As expected, increasing the port size leads to a decrease in $\beta$ as the degree of mode localization increases.}
    \label{fig:comsol_mode_localization}
\end{figure*}

In this section, we argue that achieving the same $\beta$ is inherently harder in a cavity with $V\gg\lambda^3$ than in a cavity with $V\sim \lambda^3$. 
With the same loaded $Q$ (determined by $\beta$ and $Q_0$), the steady-state electromagnetic energy density $U_{EM}$ on resonance is the same for the typical-volume cavity and the high-volume cavity. For the large-volume cavity, the Ohmic power dissipation rate on the cavity walls $\kappa_l$ is larger due to the larger surface area (Eqn. 19.150 in \cite{zangwill}).  The strength of the port $\kappa_m$ must then be proportionally larger to maintain the same coupling/Ohmic-loss ratio prescribed by $\beta=\kappa_m/\kappa_l$.  

Another way to understand this is to consider the ring down time $Q/\omega_c$, which is the same for both $V\gg\lambda^3$ and $V\sim \lambda^3$ cavities (Eqn. 19.153 in  \cite{zangwill}). To drain the much higher total energy $U_{EM}\times V$ stored in the  $V\gg\lambda^3$ cavity within the same ring-down time, the physical size of the port must be proportionally larger than the port on the typically sized cavity. 
%Another way to see this is that in a stationary scattering measurement, the ratio of the power dissipated on the metallic walls versus the reflected power 
 This also explains the proportionally larger signal strength from axion-converted photons in a high-volume haloscope at a fixed wavelength. 

In Fig. \ref{fig:couplingvera}, we illustrate these two cavities ($\sim\lambda^3$ vs $\gg\lambda^3$) using the simplest rectangular geometry.  The top panel represents a $\sim\lambda^3$ cavity with a coupling at the desired $\beta$ parameter.  The middle and bottom panels represent the $\gg\lambda^3$ cavity with proportionally larger coupling ports to maintain the same target $\beta$ (and the ring down time).  In the middle panel, it is apparent that the larger port implemented locally would severely distort the field distribution.  
%In the limit of a very long cavity the proportionally large port must eventually destroy the cavity itself. 

%\color{red}

In Fig. \ref{fig:comsol_mode_localization} (left) we demonstrate this behavior using finite element analysis (FEA) performed in COMSOL-RF. We start with a simple rectangular cavity dimensioned to resonate at 10 GHz. Its volume, $6~\textrm{L} = 221 \lambda^3$, was chosen to emulate a system with $V \gg \lambda^3$. When coupling to the cavity with a small rectangular waveguide (upper contour plot) the energy lost at the cavity walls is much higher than the energy extracted via the waveguide, leading to a low $\beta$. The mode also begins to localize on the opposite end of the cavity. When the size of the port increases in an attempt to extract more energy (lower contour plot), the degree of mode localization increases, further decreasing $\beta$. 

In Fig. \ref{fig:comsol_mode_localization} (right) we show simulated $\beta$ values for a variety of waveguide sizes. In each case, we computed $\beta$ by taking
\begin{align}
    \beta = \frac{P_{readout}}{P_{cavity~walls}} &= \frac{\omega_c \frac{U}{Q_r}}{\omega_c \frac{U}{Q_0}} = \frac{Q_0}{Q_r} \nonumber \\
&= \frac{ \int_{port~output} dA~ \vec{S}\cdot\hat{n}}{\int_{cavity~walls} dA~ \vec{S} \cdot\hat{n}},
\end{align}
where $P_{readout}$ is the power lost through the waveguide output, $P_{cavity~walls}$ is the power lost at the cavity walls, $\omega_c$ is the resonant frequency of the cavity, $U$ is the stored energy, $Q_r$ quantifies coupling to the receiver, $\vec{S}$ is the Poynting vector and $\hat{n}$ is a unit vector normal to the surface of integration. 
\begin{comment}
Since we can also write $P_{readout} = \omega_c U / Q_r$, where $\omega_c$ is the resonant frequency of the cavity, $U$ is the stored energy, and $Q_r$ characterizes coupling to the receiver, and $P_{cavity~walls} = \omega_c U / Q_0$, $\beta = Q_0/Q_r$, as expected. 
\end{comment}
Instead of extracting more power from the cavity, $\beta$ falls off rapidly for larger port sizes.

We also ran a series of similar simulations where several identically sized waveguides (scale = 1) were distributed symmetrically across the cavity volume. In these configurations, the mode remained centralized, and, by summing the port outputs, more energy could be extracted from the cavity relative to the losses at the cavity walls, increasing the coupling. The specific coupling achieved depended upon the number and geometric distribution of the waveguides. Our example cases demonstrated couplings between 0.4 and 5, representing  undercoupling, critical coupling, and overcoupling to the cavity system.

In the bottom panel of Fig. \ref{fig:couplingvera}, we imagine a readout scheme that could permit such a geometric distribution of readout ports.  An array of ports similar to the one in the top panel are distributed throughout the volume of the cavity. The signal from each port is summed coherently in a low-loss, impedance-matched tree of transmission lines.  This scheme provides a synthesized ``port" that is proportional to the volume of the cavity, in-phase for the axion signal as long as the extent of the cavity is less than the spatial coherence length of the cosmological axion (a few 100 m).

\begin{comment}
\color{red} We simulated a similar system in COMSOL-RF and achieved undercoupling, critical coupling, and overcoupling by varying the number and spatial distribution of the ports. \color{black}
\end{comment}

\begin{comment}
In Figure \ref{fig:comsol_mode_localization} (bottom), we show that a collection of waveguide ports (all Scale = 1) distributed across the cavity can achieve a variety of couplings, including undercoupling, $\sim$critical coupling, and overcoupling. This scenario is similar to the behavior predicted for the bottom panel of Fig. \ref{fig:couplingvera}.
\end{comment}

The large volume of a beehive cavity is achieved by having a large number of cells. It is easy to conclude that when the coupling of a single port required to reach a certain $\beta$ (for the entire beehive) exceeds the coupling between cells, the port will draw power only from the cell at which it is located. After reaching that point, multiple ports + signal summing will be needed to utilize the entire volume of the beehive for the axion searches. 

Such a summing tree was proposed for an axion experiment in \cite{Kuo_2020} in the context of mode selection.  Here we argue that even to maintain $\beta$ one must implement such structures. It is becoming clear that summing tree implementation should become a focus of R\&D in the VERA program. 

%\begin{widetext}

\section{Conclusion}
\label{sec:conclude}

In this paper, We provide a comprehensive comparison of linear amplifiers and microwave photon-counters in axion dark matter experiments.  We derive expressions for the scan rate of either sensing techniques, assuming a range of realistic operating conditions and detector parameters, over the frequency range between 1--30 GHz. 

As expected, photon-counting detectors such as SMPD can be overwhelmingly advantageous under low background, at high frequencies ($\nu>$ 5 GHz), {\em if} they can be implemented with appropriate parameters. We also note an expanded applicability of off-resonance photon background reduction, including the single-quadrature state squeezing, for scan rate enhancements and a much broader appeal for operating the haloscope resonators in the over-coupling regime, up to $\beta\sim 10$.

In \S\ref{sec:dfsz} we investigate the experimental reach for a wide range of scenarios.  With the goal of eventually reaching DFSZ up to 30 GHz, we identify several R\&\!D thrusts:

\begin{enumerate}
\item Develop high-volume ($\gg\lambda^3$) haloscope resonators such as VERA, the plasma haloscopes, and the dielectric disk haloscopes. A good milestone is the the reduction in the steepness of frequency scaling from $V\sim\nu^{-3}$ to $\nu^{-0.5}$. 
\item Develop SMPDs or similar microwave photon-counting detectors to have either of the following properties:
\begin{enumerate}
\item Bandwidth $\lesssim$ 0.1\% and tunable range $\gtrsim$ 20\%;
\item Bandwidth $\sim$ 20\% and very low DCR ($\lesssim$ 1 $s^{-1}$).
\end{enumerate}
\item Lower the photon temperature of the haloscope resonators and various backgrounds to $\lesssim$ 30 mK.
\end{enumerate}
   
Item (1) alone should represent a robust approach to reach KSVZ up to 10 GHz using existing sensor technology in a large-$B_0^2V$ magnet.  The combination (2b)+(3) will reach DFSZ up to 12 GHz and KSVZ to 23 GHz (Fig.~\ref{fig:compare3}).  The combinations (1)+(2a) or (1)+(2b)+(3) will cover the entire $<30$ GHz range to DFSZ (Fig.~\ref{fig:compare2}). 

\vspace{0.5 cm}

\section*{Acknowledgments}

The authors thank P. Bertet and E. Flurin for suggesting the cooling of the termination as a means to reduce photon noise, motivating overcoupling the cavity haloscope. We acknowledge C. Braggio, G. Carugno, and B. Majorovitz for useful discussions while visiting INFN and MPP. 

This material is based upon work supported by the National Science Foundation under Grant No. 2209576. We acknowledge the support of the Natural Sciences and Engineering Research Council of Canada (NSERC), 521528828. CLK, CLB, TAD, and NAK's investigation on the readout of high-volume axion haloscopes were supported by the Department of Energy, Laboratory Directed Research and Development program at SLAC National Accelerator Laboratory, under contract DE-AC02-76SF00515.  AC is supported via the Fermi National Accelerator Laboratory (Fermilab), a U.S. Department of Energy, Office of Science, Office of High Energy Physics HEP User Facility. Fermilab is managed by Fermi Research Alliance, LLC (FRA), acting under Contract No. DE-AC02-07CH11359.  CZ is supported by a KIPAC Porat Fellowship.

\bibliography{references}

\end{document}